\documentclass[twocolumn, superscriptaddress, amsmath,amssymb, aps, pra, longbibliography]{revtex4-1}

\usepackage{soul}
\usepackage{amsmath}   
\usepackage{amssymb}
\usepackage{mathtools}
\usepackage{graphicx}
\usepackage{dcolumn}
\usepackage{bm}
\usepackage{amsmath}
\usepackage{graphicx}
\usepackage{amsfonts}
\usepackage{graphicx}
\usepackage{array}
\usepackage{float}
\usepackage{ragged2e}
\usepackage{subfigure}
\usepackage{subcaption}
\usepackage{color}
\usepackage{multirow}
\usepackage{amssymb}%
\usepackage[colorlinks=true,linkcolor=blue]{hyperref}%
\hypersetup{allcolors=blue}
\usepackage[normalem]{ulem}
\usepackage{xcolor}
\usepackage{mathtools}
\usepackage[dvipsnames]{xcolor}

\newcommand{\ket}[1]{\ensuremath{\left\vert #1 \right\rangle}}
\DeclarePairedDelimiterX\braket[2]{\langle}{\rangle}{#1 \delimsize\vert #2}

\begin{document}
%\captionsetup{justification=justified, singlelinecheck=false}

\title{$^{85}$Rb  electromagnetically induced transparency and absorption on two collinear $5S_{1/2}\rightarrow5P_{1/2}\rightarrow5D_{3/2}\rightarrow 25F_{5/2}$ three-photon cascade schemes}
\date{\today }

\author{Shuaijie Li}
    \email{shuaijie@umich.edu}
    \affiliation{Department of Physics, University of Michigan, Ann Arbor, MI 48109, USA}     
\author{Alisher Duspayev}
    \thanks{Present address: Rydberg Technologies Inc., Ann Arbor, MI 48103, USA} 
    \affiliation{Department of Physics, University of Michigan, Ann Arbor, MI 48109, USA}    
\author{Georg Raithel}
    \affiliation{Department of Physics, University of Michigan, Ann Arbor, MI 48109, USA} 

\begin{abstract}
We present a comprehensive study of three-photon electromagnetically induced transparency (EIT) and absorption (EIA) in room-temperature Rb vapor. Using the cascade system $5S_{1/2}\rightarrow5P_{1/2}\rightarrow5D_{3/2}\rightarrow 25F_{5/2}$ in $^{85}$Rb, we explore two collinear laser-beam configurations. The detunings of the upper two lasers of the cascade system are varied. In both configurations, the probe transmission maps contain rich sets of EIT/EIA features and substructures. To explain these, we utilize a dressed-atom and a quantum Monte Carlo trajectory method. The latter includes 108 relevant fine- and hyperfine-structure levels and their magnetic substates, accounting for coherent optical drives, optical pumping, atom recycling, electromagnetic decays via the $5P_{J}$ levels, and velocity-changing and hyperfine collisions. We find quantum-trajectory solutions of the Lindblad master equation, which show good agreement with experimental data and aid in finding physical explanations for the observed spectra. To further explore collision-induced spectral features, collisional loss of EIT/EIA visibility, velocity-changing and hyperfine collisions, we acquire spectra in Rb cells with 50 and 500~mTorr of Ar gas.  Our work is of interest to studies of multi-level EIT systems that underpin emerging quantum technologies, where accurate accounting of internal atomic structure, laser beam geometry, optical pumping, transit time, and collisional effects is critical.
\end{abstract}

\maketitle

% ----------------------------------------------------------------
\section{Introduction}
\label{sec:intro}
Electromagnetically induced transparency (EIT)~\cite{boller1991, Scullybook} is a quantum interference effect widely employed in cold-atom-based quantum-optical experiments~\cite{EITSWM, SWM, EITSinglePhoton1}, quantum information science~\cite{Gorshkov2011, EITSinglePhoton2}, vapor-cell atomic spectroscopy~\cite{mohapatra2007, EITStarkEffect, EITQuantumDefect}, and quantum metrology~\cite{Holloway2014}. EIT occurs when a strong control field modifies the absorption properties of an atomic vapor, through which a weak probe field propagates, creating a transparency window in the probe transmission~\cite{boller1991}. EIT spectra often exhibit features caused by hyperfine structure of the utilized energy levels~\cite{Kondo:15, Duspayev2025}, laser beam effects~\cite{Ma_jphyscomm2020, tang2025}, additional inert gases in the cell~\cite{Thaicharoen2024, dash2024}, external fields~\cite{Ma2017, Das:21, DAS2024112340, PhysRevA.73.033803, PhysRevA.65.023814, schlossberger2026}, and optical pumping~\cite{Badger_2001, Yang2011, he2013, QMC2018OpticsExpress, Xue2019, sujap2022, OP_Cs, PhysRevA.73.033803, Moon:08}.  
Recent studies have highlighted the effects of hyperfine-structure-dependent collisions on EIT spectra, which require detailed modeling for explanation~\cite{Badger_2001, Duspayev2025}, 
in contrast to simulations using reduced Hilbert spaces.

Three-photon EIT~\cite{carr2012} can offer advantages in relevant applications, including all-infrared laser systems~\cite{Thaicharoen2019}, reduction of Doppler broadening~\cite{AngleMatching, bohaichuk2023}, access to high-angular-momentum Rydberg states~\cite{brown2023, prajapati2023highell, highellpaper, allinson2024}, and elimination of photo-induced stray electric fields in vapor cells~\cite{jau2020, ma20, highellpaper, patrick2025}. In both two- and three-photon EIT, introducing inert gases such as neon and argon have facilitated studies on state- and species-dependent collisional effects~\cite{Thaicharoen2024, dash2024, shieldingpaper} and electric-field sensing in plasmas~\cite{dash2025}. However, three-photon EIT usually involves intermediate states with hyperfine structure splittings ranging from a few tens to hundreds of MHz~\cite{allegrini2022}, which are large enough to cause complex observable sub-structures in experimental spectra~\cite{Duspayev2025}. Simplified numerical solutions of the Lindblad equation with up to 10 included levels, where optical pumping between the magnetic sublevels caused by the EIT lasers was neglected, had some success in explaining the observed patterns~\cite{Duspayev2025}. However, a description that captures all spectral features related to magnetic substructure, optical pumping, atom recycling, and collision-induced dynamics remains to be formulated. Results of such a comprehensive treatment will address modeling requirements in the aforementioned applications.

In the present work, we study collinear three-photon Rydberg-EIT driven on the cascade $5S_{1/2}\rightarrow5P_{1/2}\rightarrow5D_{3/2}\rightarrow 25F_{5/2}$ in $^{85}$Rb. The spectra for two configurations of the Rydberg laser beam propagation are investigated. In both cases, prominent spectral features are observed as the frequency of the laser driving the $5P_{1/2}\rightarrow5D_{3/2}$ transition is varied across the hyperfine structure of the $5D_{3/2}$ state. The comparatively large hyperfine structure of the $5P_{1/2}$ state gives rise to additional spectral features. To explain the experimental data, we develop a model in a state space that includes all hyperfine manifolds and magnetic substates coupled by the three excitation lasers and all $5P_{3/2}$ states. The latter are coupled via spontaneous decay from $5D_{3/2}$ to $5P_{3/2}$, a decay channel that accounts for about 18$\%$ of $5D_{3/2}$ decays into $5P_J$ levels.

Our experimental setup and data acquisition procedure are described in Sec.~\ref{sec:setup}. An overview of the theoretical model treated with the quantum Monte Carlo method is presented in Sec.~\ref{sec:BriefQMCWF}, with more details provided in the Supplemental Material~\cite{SM}.
%Supplementary Information. 
Experimental results for Rydberg-EIT in Rb vapor cells and corresponding simulations are presented and discussed in Sec.~\ref{sec:experimentalResults}. These results are further analyzed in Sec.~\ref{sec:model} using a dressed-atom picture. Finally, results for three-photon Rydberg-EIT in Rb vapor cells with 50~mTorr and 500~mTorr of argon are analyzed in Sec.~\ref{sec:experimentalResultsWithAr}. The paper is concluded in Sec.~\ref{sec:conclusion}.

\section{Experimental setup}
\label{sec:setup}
\subsection{Three-photon Rydberg-EIT}
\label{sec:excitation scheme}

\begin{figure}[t!]
  \centering
  \includegraphics[width=0.48\textwidth]{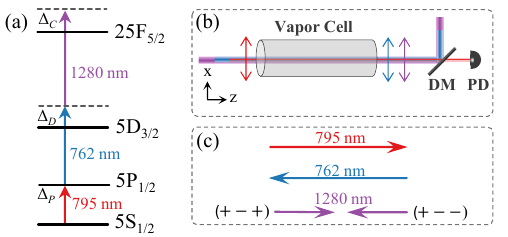}
  \caption{\justifying (a) Energy level diagram with laser wavelengths and detunings $\Delta_i~(i = P, D, C)$ of the probe, dressing and coupler lasers, respectively. (b) Experimental layout. Lasers are linearly polarized along $x$. DM and PD denote a %bandpass 
  dichroic mirror and a photodiode, respectively. (c) Illustration of the $(+-+)$ and $(+--)$ configurations, defined by beam directions relative to the $z$-axis.}
  \label{fig:3p excitation}
\end{figure}

The three-photon Rydberg excitation scheme, illustrated in Fig.~\ref{fig:3p excitation}(a), involves three collinear infrared lasers overlapped with each other. The probe laser is at 795~nm and is resonant with the $\ket{5S_{1/2}, F=3} \rightarrow \ket{5P_{1/2}, F^{\prime}=3}$ transition (detuning $\Delta_P=0$). The dressing laser is at 762~nm and is tuned across the $\ket{5P_{1/2}, F^{\prime}=3} \rightarrow \ket{5D_{3/2}, F^{\prime\prime}}$ hyperfine transitions with detuning $\Delta_D$ measured relative to the $F'=3$ to $F''=4$ component. The coupler (Rydberg) laser is at 1280~nm and is scanned across the $\ket{5D_{3/2}, F^{\prime\prime}} \rightarrow \ket{25F_{5/2}, F^{\prime\prime\prime}}$ transitions. The hyperfine structure of $25F_{5/2}$, estimated to be on the order of $h \times 1$~kHz based on the $n^{-3}$-scaling of hyperfine splittings with the principal quantum number $n$~\cite{gallagher}, is unresolved and negligible. The coupler detuning $\Delta_C$ is measured relative to the $\ket{5D_{3/2}, F^{\prime\prime} = 4}\rightarrow \ket{25F_{5/2}, F'''=*}$ transition.

The probe and dressing beams propagate along the $+z$ and $-z$-directions, respectively [see Fig.~\ref{fig:3p excitation}(c)]. The propagation direction of the coupler beam is either in the $+z$ or $-z$ direction, corresponding to configurations denoted as $(+-+)$ or $(+--)$, respectively. The probe beam passes through the 7.5-cm-long room-temperature vapor cell and a 
dichroic mirror (DM), which reflects the dressing and coupler beams while transmitting the probe beam. The latter is then detected by a photodiode (PD).

The probe beam is focused to a waist of $w_0 \approx 40~\mu\mathrm{m}$ and has a power of $\approx8~\mu\mathrm{W}$. The dressing and coupler beams have waists of $\approx 100~\mu\mathrm{m}$ and $\approx 170~\mu\mathrm{m}$ 
and powers of $\approx 1~\mathrm{mW}$ and $\approx 10~\mathrm{mW}$, respectively. For these parameters, the radial Rabi frequencies at the beam centers are  $\Omega_{PS} \approx 100~\mathrm{MHz}$ for $5S_{1/2} \rightarrow 5P_{1/2}$, $\Omega_{DP} \approx 120~\mathrm{MHz}$ for $5P_{1/2} \rightarrow 5D_{3/2}$, and $\Omega_{CD} \approx 30~\mathrm{MHz}$ for $5D_{3/2} \rightarrow 25F_{5/2}$.

\subsection{Data acquisition system}
\label{sec:DAQ}

To increase the signal-to-noise ratio (SNR) of the EIT signal, we amplitude-modulate the coupler beam using an acousto-optic modulator [not shown in Fig.~\ref{fig:3p excitation}(b)] with a square-pulse repetition rate of 1.5~kHz and a duty cycle of 50\%. 
The PD signal is demodulated using a lock-in amplifier with an output time constant of 3~ms.

The signal $S_0(\Delta_C, \Delta_D)$ is acquired on an oscilloscope in ac-coupled mode, as $\Delta_C$ is scanned over several hundreds of~MHz, yielding  two-dimensional spectral maps on the $(\Delta_C, \Delta_D)$-plane, 
\begin{equation}
S(\Delta_C, \Delta_D)=S_0(\Delta_C, \Delta_D)-\langle S_0(\Delta_C, \Delta_D)\rangle_{\Delta_C}\quad.
\label{eq:sdef}
\end{equation}
For each
$\Delta_D$-value,
$S_0(\Delta_C, \Delta_D)$ is averaged over several hundred coupler-laser scans, with $\Delta_D$ held fixed and for an individual-scan duration of 100~ms.
For each $\Delta_D$-value, we subtract the $\Delta_C$-average, $\langle S_0(\Delta_C, \Delta_D)\rangle_{\Delta_C}$, to obtain $S(\Delta_C, \Delta_D)$. This ac-coupling method removes the two-photon EIT effects attributed to the $5S_{1/2}\rightarrow5P_{1/2}\rightarrow5D_{3/2}$ ladder cascade, while retaining the three-photon Rydberg-EIT and electromagnetically induced absorption (EIA) effects we are interested in.

The dressing laser is offset-locked to the $\ket{5P_{1/2}, F^{\prime}=3} \rightarrow \ket{5D_{3/2}, F''=4}$ transition, allowing us to vary the detuning $\Delta_D$ in well-defined 2-MHz steps from $-60$ to $+20$~MHz. This range, together with the $\Delta_C$ scan range of $-450$~MHz$\, \lesssim \Delta_C \lesssim \,450$~MHz, is sufficient to capture the major features in the EIT spectra.

To calibrate the coupler-laser frequency scan, a beam sample is sent through an electro-optic modulator driven at 125~MHz. The modulated beam is transmitted through a Fabry--P\'erot (FP) etalon with a free spectral range of 375~MHz. As the coupler laser is scanned, the FP has transmission peaks separated by 125~MHz. These are recorded simultaneously with the EIT spectra and are used to calibrate the $\Delta_C$-axis.

\section{Quantum Monte Carlo Wave-Function Simulation}
\label{sec:BriefQMCWF}

To model the observed three-photon EIT spectra, we simulate the open-system dynamics of the multi-level $^{85}$Rb atom using a quantum Monte Carlo wave-function (QMCWF) method, which provides a stochastic solution of the time-dependent Lindblad equation. The simulation includes all relevant hyperfine and Zeeman sub-levels of the $5S_{1/2}$, $5P_{1/2}$, $5D_{3/2}$, $25F_{5/2}$, and $5P_{3/2}$ states, resulting in a Hilbert space of 108 states. These are coupled by coherent optical drives, spontaneous decay, and optional collisional hyperfine mixing.
The $5P_{3/2}$ states, which become populated through spontaneous decay of atoms in $5D_{3/2}$, influence optical pumping and state distribution of the atoms.

The system dynamics are obtained from ensembles of typically 500 quantum trajectories for one set of parameters $(\Delta_C, \Delta_D, v_z)$, with atom velocity $v_z$ in $z$-direction. Each trajectory consists of intervals of non-Hermitian Hamiltonian evolution interrupted by stochastic quantum jumps. Atomic center-of-mass motion is incorporated by averaging over Maxwell-distributed velocities $v_z$ and using randomly selected atom-field interaction times determined under the assumption of circular top-hat optical beam profiles.   
The probe absorption spectra, quantified by the probe's Beer's absorption coefficient, are derived from the density operator, $\hat{\rho}$, which is constructed from an average over the quantum trajectory ensembles. 
The method is suitable for treating systems with large Hilbert spaces~\cite{Xue2019, QMC2018OpticsExpress}, such as ours. In the simulated data shown below, we display the ac-coupled Beer's absorption coefficient, $\Delta \tilde{\alpha} (\Delta_C, \Delta_D)  = \alpha (\Delta_C, \Delta_D) -
\langle \alpha (\Delta_C, \Delta_D) \rangle_{\Delta_C}$ with $\Delta_D$ held fixed in the $\langle \rangle_{\Delta_C}$-average. The absorption coefficient $\alpha$ is given by
\begin{widetext}
\begin{equation}
\alpha (\Delta_C, \Delta_D)  = 
\frac {4 \pi \, n_{\rm{V}}} {h \epsilon_0 \lambda_P}
\sum_{F', m_{F}} 
\frac {\int 
 P_{v_z} (v_z)\,(d_{F=3,m_F}^{F', m_{F}})^2 \,{\rm{Im}}[\rho_{F', m_F}^{  F=3,m_F}] d v_z} {\Omega_{F=3,m_F}^{F',m_{F}}}\quad.
\label{eq:alpha}
\end{equation}
\end{widetext}
There, $P_{v_z}$ is the one-dimensional Maxwell velocity distribution, $\lambda_P = 795$~nm is the probe-laser wavelength, 
$\rho_{F', m_F}^{  F=3,m_F}$ is the density matrix element $\langle 5S_{1/2}, F=3, m_F \vert \hat{\rho}(\Delta_C, \Delta_D, v_z) \vert 5P_{1/2}, F',m_{F}  \rangle$, 
$\Omega_{F=3,m_F}^{F',m_{F}}$ is the Rabi frequency (in $\mathrm{Hz}$) of the probe transition between those states for our $\pi$-polarized probe, and $d_{F=3,m_F}^{F',m_{F}}$ is the corresponding transition dipole moment in ($\mathrm{C}~\mathrm{m}$). For the internal dynamics, we use the $x$-direction as the quantization axis because all lasers are $x$-polarized (see Fig.~\ref{fig:3p excitation}). A detailed description of the model, including Hilbert-space setup, coherent couplings, decay channels, numerical implementation, and parameters used in the simulations, is provided in the Supplemental Material~\cite{SM}.

By virtue of utilizing a Hilbert space that incorporates all hyperfine levels and their magnetic substates,  the QMCWF model 
covers EIT, EIA and optical pumping, which exhibit significant cross-talk.
To make the simulation feasible, we neglect some excited states, including $5D_{5/2}$, $6P_J$, and $25F_{7/2}$. These excited states are far-off-resonant, are not allowed for the given laser drives, or become populated at comparatively low rates.
Further, radiative decay rates of the Rydberg states are very low compared with the inverse of the atom-field interaction time ($\sim 10^6~\mathrm{s^{-1}}$), rendering such decays unimportant relative to kinetic atom recycling. As our results will show, the 108-state simulation exhibits a remarkable degree of detailed agreement with corresponding experimental data. 
To aid with the interpretation of the observed EIT modes, we will also use a simplified dressed-atom model.

\section{Experimental Results for vapor cells without added background gas}

\label{sec:experimentalResults}
 
In this section, we present experimental results and corresponding QMCWF simulations for the $(+-+)$ and $(+--)$ configurations in Rb vapor cells without added background gas. The discussed spectral features are interpreted based on results from the QMCWF simulations and the dressed-atom model introduced in Sec.~\ref{sec:model}.

\begin{figure}[htb]
  \centering
  \includegraphics[width=0.46\textwidth]{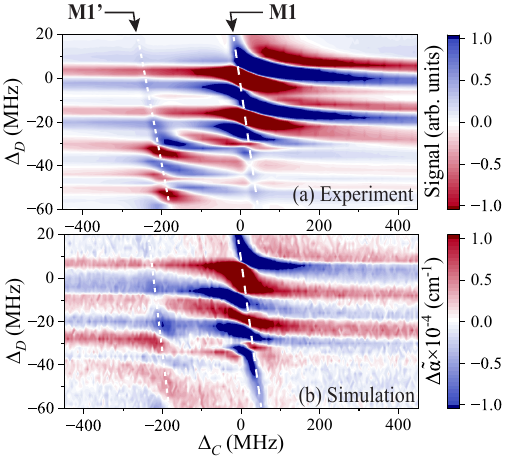}
  \caption{\justifying{Results for the $(+--)$ configuration. 
  (a) Experimental ac-coupled signals $S(\Delta_C, \Delta_D)$ [see Eq.~(\ref{eq:sdef})] from Rb cells without Ar buffer gas.
  (b) Corresponding QMCWF simulation, showing the change in the probe absorption coefficient, $\Delta \tilde{\alpha} (\Delta_C, \Delta_D)$ [see Eq.~(\ref{eq:alpha2})]. The number of quantum trajectories for each set of parameters $(\Delta_C, \Delta_D, v_z)$ is 500. In both (a) and (b), data are displayed on linear color scales; data values outside the color-bar range are shown in dark blue or dark red. Blue (red) regions correspond to three-photon EIT (EIA). White dashed lines trace out the signal modes M1 and M1' (see text).}}
  \label{fig:Rb+--} 
\end{figure}

\subsection{$(+--)$ configuration}
\label{subsec:Rb_cell_+--}

Fig.~\ref{fig:Rb+--}(a) shows an experimental map of 
$S(\Delta_C, \Delta_D)$, as defined in Eq.~(\ref{eq:sdef}),
in the $(+--)$ configuration. In the color map, blue and red regions correspond to EIT and EIA signals, respectively. 
Corresponding QMCWF simulation results are presented in Fig.~\ref{fig:Rb+--}(b), where we plot the change in probe absorption coefficient, $\Delta\tilde{\alpha}$, relative to the average $\langle \alpha(\Delta_C, \Delta_D) \rangle_{\Delta_C}$. Note that increasing (decreasing) $\Delta\tilde{\alpha}$ corresponds to red (blue) regions to match the EIA (EIT) patterns in Fig.~\ref{fig:Rb+--}(a).

Figs.~\ref{fig:Rb+--}(a) and~\ref{fig:Rb+--}(b) exhibit alternating EIT and EIA features that are lined up along diagonals, denoted as  modes M1 and M1'. These features extend to the edges of the displayed range $-60~\mathrm{MHz}\lesssim\Delta_D\lesssim20~\mathrm{MHz}$, and they are located at $\Delta_C \approx 20$~MHz and $\Delta_C \approx - 220$~MHz, respectively. The M1 and M1' features arise from 
atoms subject to resonant single-photon excitation by the probe laser on the transitions $\ket{5S_{1/2},F=3}\rightarrow\ket{5P_{1/2},F'}$, and two-photon Rydberg-state excitation $\ket{5P_{1/2},F'}\rightarrow\ket{5D_{3/2},F''}\rightarrow\ket{25F_{5/2},F'''}$, with $F'=3$ for M1 and $F'=2$ for M1', and $F''$ ranging from $F'-1$ to $F'+1$, and $F'''$ ranging from $F''-1$ to $F''+1$. Given that the 5$P_{1/2}$ hyperfine splitting is 362~MHz~\cite{SteckRb85} and that the probe detuning is $\Delta_P\approx 0$, in the modes M1 and M1' the probe predominantly excites atoms with longitudinal velocities $v_z\approx 0$ and $v_z\approx 290~\mathrm{m/s}$, respectively. Due to the low Maxwell-distribution probability at $290~\mathrm{m/s}$, the M1' feature is relatively weak compared to M1. Also, both M1 and M1' features follow trend lines with identical slopes $\delta \Delta_D/\delta \Delta_C \approx -1.0$. This trend is due to the requirement of $5S_{1/2} \rightarrow 25F_{5/2}$ three-photon resonance in cases of low saturation and large dressing- and coupler-beam detunings from intermediate hyperfine resonances. Further, M1' is shifted relative to M1 by $-240$~MHz in $\Delta_C$ due to dressing- and coupler-beam Doppler shifts and the 5$P_{1/2}$ hyperfine splitting. Near the perimeters of the displayed $\Delta_D$-range, in most cases, the intermediate hyperfine levels of $5D_{3/2}$ are sufficiently far-off-resonant to avoid significant population and photon scattering from these levels. This results in fairly narrow M1 EIT lines at both $\Delta_D$-perimeters, and in a fairly narrow M1' line at the upper $\Delta_D$-perimeter.
These remain visible even at fairly large $\Delta_D$ because the drop-off of the two-photon transition matrix elements from $5P_{1/2}$ to $25F_{5/2}$ as a function of intermediate detunings from the $5D_{3/2}$ levels is quite slow, as the two-photon matrix elements scale inversely with intermediate detuning.

As the dressing detuning $\Delta_D$ is varied, $5P_{1/2}$-atoms, which predominantly have velocities $v_z \approx 0$ or $\approx 290$~m/s, become resonantly excited into hyperfine levels of $5D_{3/2}$ at certain $\Delta_D$ values. This  leads to prominent horizontal bars and anti-crossings intersecting the M1 and M1' features in Fig.~\ref{fig:Rb+--}. For instance, 
the anti-crossings with M1 are near locations $(\Delta_C=-\delta_{5D_{3/2},F''},\, \Delta_D = \delta_{5D_{3/2},F''})$ with  relative hyperfine energies %relative to $F''=4$, namely 
$\delta_{5D_{3/2},F''}$ $\approx0$, $-18~\mathrm{MHz}$, and $-30~\mathrm{MHz}$ for $F''=4$, 3 and 2, respectively~\cite{nez1993}.

The described properties of the M1 and M1' features agree well between experiment and simulations.
%and dressed-atom models.
Fig.~\ref{fig:Rb+--}(b) shows results for the simulated ac-coupled absorption coefficient, 
\begin{equation}
\Delta \tilde{\alpha} (\Delta_C, \Delta_D) = 
\alpha(\Delta_C, \Delta_D) - \langle \alpha(\Delta_C, \Delta_D) \rangle_{\Delta_C}\quad,
\label{eq:alpha2}
\end{equation}
with $\alpha (\Delta_C, \Delta_D)$ from Eq.~(\ref{eq:alpha}).
In the simulation in Fig.~\ref{fig:Rb+--}(b), we use top-hat beam profiles with a radius of 40~$\mu$m, radial Rabi frequencies of 70~MHz, 80~MHz, and 20~MHz for the probe, dressing, and coupler lasers, respectively, and atomic data for $^{85}$Rb~\cite{SteckRb85, nez1993}. 
These and other simulations show that 
specific details of the simulation results depend on the Rabi frequencies, as expected. Also, the three-photon cascade system considered is open to decay into the uncoupled ground state $\ket{5S_{1/2}, F=2}$, which is included in our model. In the absence of ground-state hyperfine repumping, having a small laser beam radius is important because it limits the atom-field interaction time to $\lesssim 1~\mu$s, preventing overwhelming decay of the atoms into the uncoupled state $\ket{5S_{1/2}, F=2}$.

\subsection{$(+-+)$ configuration}
\label{subsec:Rb_cell_+-+}

\begin{figure}[htb]
  \centering  \includegraphics[width=0.5\textwidth]{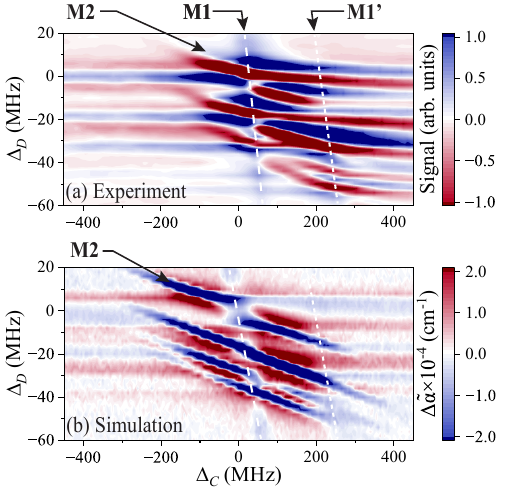}
  \caption{\justifying{Results for the $(+-+)$ configuration. The plots are organized as in} Fig.~\ref{fig:Rb+--}.}
  \label{fig:Rb+-+}
\end{figure}

Next, we investigate the $(+-+)$ configuration by reversing the coupler laser's propagation direction. The corresponding results for $S(\Delta_C, \Delta_D)$ and $\Delta \tilde{\alpha} (\Delta_C, \Delta_D)$ are shown in Figs.~\ref{fig:Rb+-+}(a) and~\ref{fig:Rb+-+}(b), respectively. 

Similar to the discussion in Sec.~\ref{subsec:Rb_cell_+--}, resonant excitation of atoms moving at $v_z \approx 0$ and $v_z\approx 290~\mathrm{m/s}$ through $|5P_{1/2}, F'=3\rangle$ and $|5P_{1/2}, F'=2\rangle$, respectively, and two-photon Rydberg excitation, lead to M1 and M1' modes
with slopes of $\delta\Delta_D/\delta\Delta_C \approx -1.0$. Here, the M1 and M1' modes appear as rather faint, localized variations imprinted onto the dominant M2 features discussed below. Moreover, the M1' mode is shifted in $\Delta_c$ by $+209$~MHz from the M1 mode, as the sign of the coupler beam's Doppler shift is reversed.

A comparison of Figs.~\ref{fig:Rb+--}(a) and~\ref{fig:Rb+-+}(a) shows that the M1 and M1' signals in the $(+--)$ configuration are narrower and much stronger than in the $(+-+)$ configuration. 
Experimentally, we find that the EIT peak amplitudes at large $|\Delta_D|$ in the $(+--)$ configuration are up to approximately four times as large as those in the $(+-+)$ configuration. This enhancement 
generally favors the $(+--)$ three-photon Rydberg-EIT configuration over the $(+-+)$ one in applications where strong signal and high SNR are required~\cite{Thaicharoen2019,brown2023,bohaichuk2023,allinson2024}.

Fig.~\ref{fig:Rb+-+} exhibits a prominent set of diagonal features, labeled M2, which emerge near the resonance points of the $|5D_{3/2}, F''\rangle$ hyperfine levels with the dressing field. 
The M2 diagonals mostly appear as interlaced EIT and EIA stripes, which are shallower 
than the modes M1 and M1' in the $(\Delta_C,\Delta_D)$ plane. In the M2 modes, the probe and dressing lasers drive a two-photon hyperfine transition $\ket{5S_{1/2},F}\rightarrow \ket{5D_{3/2},F''}$ for atoms with a velocity $v_z$ at which this transition is two-photon-resonant, while the coupler laser introduces a resonant 
single-photon coupling from $\ket{5D_{3/2}, F''}$ to the Rydberg states $\ket{25F_{5/2},F'''=*}$.
Taking Doppler shifts into account, we find that the lower-stage two-photon process selects atoms with longitudinal velocity $v_z = -18.36\,(\Delta_D - \delta_{5D_{3/2},F''})$~m/s/MHz. Considering the resultant shifts of
the upper-stage single-photon process, one finds that the M2 structures have slopes of $\delta \Delta_D/\delta \Delta_C \approx -0.070$. Due to the limited width of the Maxwell velocity distribution, the range of the M2 signals along the $\Delta_D$-axis is only on the order of 20~MHz for each $\vert 5D_{3/2}, F'' \rangle$-resonance. This is in contrast with the M1 and M1' modes, which are associated with atoms of approximately fixed velocities of $\approx 0$ and $\approx 290$~m/s. As a result, the M1 and M1' modes extend across the full range of $\Delta_D$, in both Figs.~\ref{fig:Rb+--} and~\ref{fig:Rb+-+}.
Also, in the $(+--)$ configuration in Fig.~\ref{fig:Rb+--}, equivalents of the M2 signals of Fig.~\ref{fig:Rb+-+} would have slopes of $\delta \Delta_D/\delta\Delta_C = +0.070$. Such signals are not observed in Fig.~\ref{fig:Rb+--} because they wash out when integrating the atomic absorption over $v_z$ (see discussion in Sec.~\ref{sec:model}).

Similar to the case of $(+--)$ configuration discussed in Sec.~\ref{subsec:Rb_cell_+--}, in the $(+-+)$ configuration, atoms with $v_z \approx 0$ or $\approx 290$~m/s 
become resonantly excited from $|5S_{1/2}, F=3\rangle$ into $|5P_{1/2}, F'=3\rangle$ or $|5P_{1/2}, F'=2\rangle$, respectively. Subsequent excitation into $5D_{3/2}$ leads to a largely $\Delta_C$-independent horizontal striping pattern that reflects the $5D_{3/2}$ hyperfine splittings~\cite{nez1993}.

The simulation in Fig.~\ref{fig:Rb+-+}(b) reproduces Fig.~\ref{fig:Rb+-+}(a)  moderately well. Deviations may be attributable to velocity-changing collisions with an inert background gas that may be present in the cell~\cite{Lei2025}, such as atmospheric He diffusing through the cell walls.
Velocity-changing collisions would effectively introduce ground-state hyperfine repumping. Although we have not used a repumper laser~\cite{eitrepumper2021} in the experiments, ground-state hyperfine repumping
has been added in the simulations in Fig.~\ref{fig:Rb+-+}(b) to improve agreement with the experiment. 
The ground-state hyperfine repumping of $10^6$~s$^{-1}$ applied in Fig.~\ref{fig:Rb+-+}(b) is very small in comparison with the ground-state repumping rates considered in Sec.~\ref{sec:experimentalResultsWithAr}, yet it is sufficient to enhance the visibility of the M1 and M1' signals and to bring them into closer agreement with the experiment. In the absence of repumping, optical pumping into $|5S_{1/2}, F=2\rangle$ bleaches the simulated M1 and M1' signals relative to the M2 signals beyond recognition.
Deviations between experiments and simulations may also result from intensity variations within the near-Gaussian laser beam profiles, which we have approximated as top-hat profiles in the simulations.
Future efforts to elucidate these details further will require refined simulations, possibly also including environmental magnetic fields. Regardless, the present agreement in Figs.~\ref{fig:Rb+--} and~\ref{fig:Rb+-+} between experimental and simulated maps of the complete 108-level system is remarkable.

\section{Dressed atom model}
\label{sec:model}

Next, we discuss the results presented in Sec.~\ref{sec:experimentalResults} in context with velocity-resolved Lindblad simulation data, $[d\alpha/dv_z] (\Delta_C, \Delta_D, v_z)$ and corresponding dressed-state calculations. While in the experiment we can only observe the net absorption, $\alpha(\Delta_C, \Delta_D) = \int [d\alpha/dv_z] (\Delta_C, \Delta_D, v_z) dv_z$,
the velocity-resolved simulation data add insight into the origin of the experimentally observed M1- and M2-type signals. 
For brevity, we focus on two selected values of $\Delta_D$. Also, in this section we use reduced level spaces (10 levels for the Lindblad and 3 for the dressed-state calculations) to reduce model complexity and to enhance clarity.

The $[d\alpha/dv_z] (\Delta_C, \Delta_D, v_z)$ data are obtained by integrating the Lindblad equation in the reduced 10-level state spaces $\{ \vert 5S_{1/2}, F=3 \rangle, \vert 5P_{1/2}, F'=3 \rangle, \vert 5D_{3/2} F''=2-4\rangle, \vert 25F_{5/2}, F'''=1-5 \rangle \}$ with fixed $m_F$~\cite{Duspayev2025}, before averaging over $m_F$.

Atoms moving at velocity $v_z$ experience Doppler shifts
in the laser beams, which all propagate along $z$ or $-z$.
The frequency detunings $\Delta_i$ in the field picture and in the frame of reference of moving atoms, with $i=P$, $D$, $C$ for probe, dressing, and coupler fields, respectively, are given by
\begin{equation}
\nonumber
\begin{aligned}
    &\Delta_P=\Delta_{P,0}-v_z / \lambda_P \quad, \\
    &\Delta_D=\Delta_{D,0}+v_z / \lambda_D \quad, \\
    &\Delta_C=\Delta_{C,0}\pm  v_z / \lambda_C \quad,
\end{aligned}
\end{equation}
in units of Hz. Doppler shifts of $\pm v_z / \lambda_C $ correspond to the configurations $(+-\mp)$, and the $\Delta_{i,0}$ are frequency detunings without Doppler effect.
Assuming a weak probe (probe Rabi frequency less than the upper-level decay rates and the optical Rabi frequencies), we consider the three upper dressed states that correspond to the atomic levels $\{5P_{1/2}, 5D_{3/2}, 25F_{5/2} \}$. The drive Rabi frequencies in units of Hz are $\Omega_{DP}$ and $\Omega_{CD}$ for the dressing and coupler beams, respectively, and the corresponding field-atom detunings are $\Delta_D$ and $\Delta_C$. 
The Hamiltonian in the upper 3-level system is 
\begin{widetext}
\begin{equation}
\hat{H}_{up3} = h \begin{pmatrix}
v_z / \lambda_P & \Omega_{DP}/2 & 0 \\
\Omega_{DP}^*/2 & -\Delta_D +v_z (1/\lambda_P-1/\lambda_D) & \Omega_{CD}/2 \\
0 & \Omega_{CD}^*/2 & -\Delta_D - \Delta_C +v_z (1/\lambda_P-1/\lambda_D \pm 1/\lambda_C)
\end{pmatrix}\quad.
\label{eq:hamup3}
\end{equation}
\end{widetext}
For the case $\Delta_P=0$, absorption may occur if an eigenvalue of $\hat{H}_{up3}(v_z)$ is zero, because then the corresponding eigenstate is in resonance with $5S_{1/2}$~\cite{Thaicharoen2019}. For given Rabi frequencies $\Omega_{DP}$ and $\Omega_{CD}$ and a fixed detuning $\Delta_D$ in $\hat{H}_{up3}$, there are up to three velocities $v_{z,i}$ for which Eq.~(\ref{eq:hamup3}) has a zero eigenvalue and absorption can occur.  The velocities $v_{z,i}$ are determined by finding the roots of the
third-order polynomial $\det [\hat{H}_{up3}(v_z) ] =0$. The corresponding normalized eigenstates have amplitudes $c_{5P,i}$ and $c_{5D,i}$ in the $5P_{1/2}$ and $5D_{3/2}$ base kets, respectively. 
The quantity $A_i =  ( |c_{5P,i}|^2 + |c_{5D,i}|^2 ) P_{v_z}(v_{z,i})$, with Maxwell probability $P_{v_z}(v_{z,i})$, serves as a relative measure for one- or two-photon probe-light absorption.

\begin{figure}[htb]
 \centering
  \includegraphics[width=0.5\textwidth]
{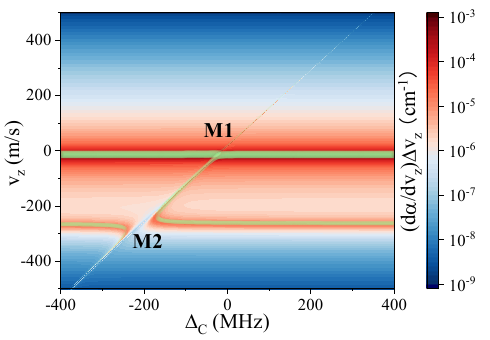}
  \caption{\justifying Velocity map for a 10-level system in the $(+-+)$ configuration. The color scale shows the calculated response, $(d\alpha/dv_z)\Delta v_z$, as a function of the coupler detuning $\Delta_C$ and the atomic velocity $v_z$. 
  The calculation is performed for $\Delta_D=15$~MHz and radial Rabi frequencies of $\Omega_{PS}=25$ MHz, $\Omega_{DP}=80$ MHz, and $\Omega_{CD}=20$ MHz.
The overlaid green symbols and their areas show the dressed-state locations $(\Delta_C,v_{z,i})$ and the corresponding qualitative absorption strengths $A_i$ in the weak-probe limit for net Rabi frequencies of
$\Omega_{DP}=30$~MHz and $\Omega_{CD}=10$~MHz. The labeled features M1 and M2 correspond to the two dominant velocity-dependent EIT branches discussed in the text.
} 
  \label{fig:figA}
\end{figure}

In Fig.~\ref{fig:figA}, we overlay the data points $(\Delta_C, v_{z,i})$ with symbol areas $\propto A_i$ over the simulated absorption maps, $[d \alpha / dv_z] (\Delta_C, \Delta_D, v_z)$, for constant $\Delta_D = 15~$MHz and in $(+-+)$ configuration. We observe two absorption bars at fixed $v_{z,i}$, interrupted by anti-crossings with a diagonal line. 
The top horizontal bar at $v_z \sim 0$ corresponds to strong absorption due to (near-)resonant excitation into the dressed state with predominantly $5P_{1/2}$ character. The top bar is slightly ac-shifted from zero velocity to $v_z \approx - \lambda_P \Omega_{DP}^2 / (4\Delta_D) = - 12$~m/s, where the Doppler shift of the probe compensates the ac shift of the level $5P_{1/2}$ caused by the dressing beam. The bottom bar at approximately $v_{z0} = \Delta_D /(1 / \lambda_D - 1/ \lambda_P)  = - \Delta_D \times 18.36$m/s/MHz $=-275$~m/s is due to absorption on the $5S_{1/2} \rightarrow 5D_{3/2}$ two-photon resonance.  The velocity of the lower bar is slightly ac-shifted from $v_{z} = v_{z0}$ to $v_{z} \approx v_{z0} + \lambda_P \Omega_{DP}^2 / (4 \Delta_D) = -263$~m/s.
% amount $\approx \lambda_P \Omega_D^2 / (-v_{z0} (1 / \lambda_D - 1/ \lambda_P) )$. 

One of the solutions of $\det (\hat{H}_{up3})=0$ asymptotically approaches a
diagonal line in the $(\Delta_C, v_z)$-plane.
In most regions of the plane, the diagonal line corresponds to a dressed state of $25F_{5/2}$-character near the three-photon resonance $5S_{1/2}$ $\rightarrow$ $25F_{5/2}$, which follows 
\[ v_z = \frac{\Delta_D + \Delta_C}{1/\lambda_P-1/\lambda_D+1/\lambda_C}\quad.\]

In the $(+-+)$ configuration, at locations away from the intersections with the horizontal absorption bars, the diagonal line has a slope $(1/\lambda_P-1/\lambda_D+1/\lambda_C)^{-1} > 0$. 
The diagonal line does not appear in the probe absorption except when it intersects with one of the absorption bars at near-constant $v_z$. The anti-crossing at $(\Delta_C \approx 0,  v_z \approx 0)$ generates EIT on the single-photon probe transition $5S_{1/2} \rightarrow 5P_{1/2}$. In this case, the resonant two-photon transition $5P_{1/2} \rightarrow 25F_{5/2}$ acts as an effective single-stage EIT coupling transition. Hence, that anti-crossing corresponds to mode M1, and its coupling strength, the two-photon Rabi frequency from $5P_{1/2}$ to $25F_{5/2}$, is $\Omega_{DP} \Omega_{CD}/ (2 \Delta_D)$. The M1 feature in the $(+-+)$ configuration 
in the observable absorption spectrum, $\alpha(\Delta_C, \Delta_D)$,
is comparatively weak because the M1 anti-crossing in Fig.~\ref{fig:figA} mostly washes out upon integration over $v_z$ due to the unfavorable sloping of the diagonal dressed-level feature.

At the intersection of the diagonal line with the bar near velocity $v_{z0}$, an anti-crossing between $5D_{3/2}$ and $25F_{5/2}$ occurs, with a width $\sim \Omega_{CD}$. As a result, the bar near velocity $v_{z0}$ breaks up and exhibits an absorption-free gap. The absorption gap remains open upon integration over $v_z$ due to favorable sloping of the diagonal dressed-level feature. The resulting strong EIT signal at $\Delta_C \approx v_{z0}/\lambda_C$ corresponds with the M2-type signals in the spectroscopic data in Fig.~\ref{fig:Rb+-+} (note $v_{z0}<0$).

\begin{figure}[t!]
 \centering
  \includegraphics[width=0.5\textwidth]
{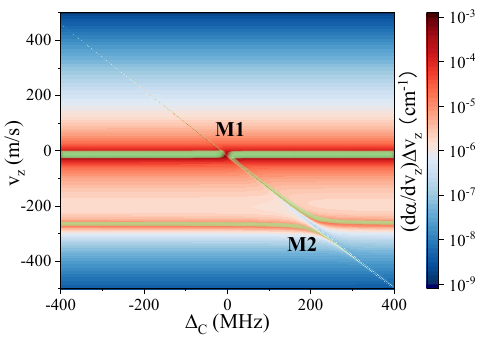}
  \caption{\justifying Velocity map and dressed states in the $(+--)$ configuration. The contents are analogous to those in Fig.~\ref{fig:figA}.
} 
  \label{fig:figAPMM}
\end{figure}

We consider the $(+--)$ case in an analogous manner in Fig.~\ref{fig:figAPMM}. The main difference is the slope of the diagonal line, which is 
$(1/\lambda_P-1/\lambda_D-1/\lambda_C)^{-1} < 0$ in the $(+--)$ case. 
As a result of the negative slope, the absorption gap at the M2 anti-crossing largely vanishes upon integration along $v_z$, suppressing the M2 signal in $(+--)$ measurements. The M1 mode is more prominent in $(+--)$ than in $(+-+)$ measurements because in the $(+--)$ case
the absorption gap at the M1 anti-crossing remains open after integration along $v_z$.

\section{Effects of background gas on EIT spectra}
\label{sec:experimentalResultsWithAr}

After explaining the physical reasons for the M1 and M2 features, we consider collisional effects due to an inert gas in the cell, which lead to modifications in the EIT spectra~\cite{Thaicharoen2024, Duspayev2025}. Relevant collision rates are estimated as follows. The rate of velocity-changing collisions of Rb in 500~mTorr of Ar is  $\gtrsim 10^7$~s$^{-1}$~\cite{Thornton2011}. Atoms that are already inside the EIT laser beams, but that are initially far-off-resonant from any atom-field interaction, are brought into resonance as a result of velocity-changing collisions inside the beams. Hence, the atom population within the atom-field interaction region experiences hyperfine repumping at a rate 
given by the rate of the velocity-changing collisions. The process amounts to atom recycling within the volume of the EIT laser beams. Notably, ground-state hyperfine and Zeeman mixing rates are negligible in comparison with the rate of velocity-changing collisions.
For $5P_{1/2}$, the rate of velocity-changing collisions is similar to that of $5S_{1/2}$. In addition, $5P_{1/2}$ hyperfine mixing occurs at significant rates near $\sim 6 \times 10^6$~s$^{-1}$~\cite{Bulos1971NuclearSpin}, while fine-structure mixing is negligible~\cite{Gallagher1968Rubidium}.
For $5D_{3/2}$, the rate of velocity-changing collisions, hyperfine, Zeeman and fine-structure mixing are all large and estimated at $\sim 2 \times 10^7$~s$^{-1}$
~\cite{Duspayev2025}. 

\begin{figure}[t]
    \centering    \includegraphics[width=0.5\textwidth]{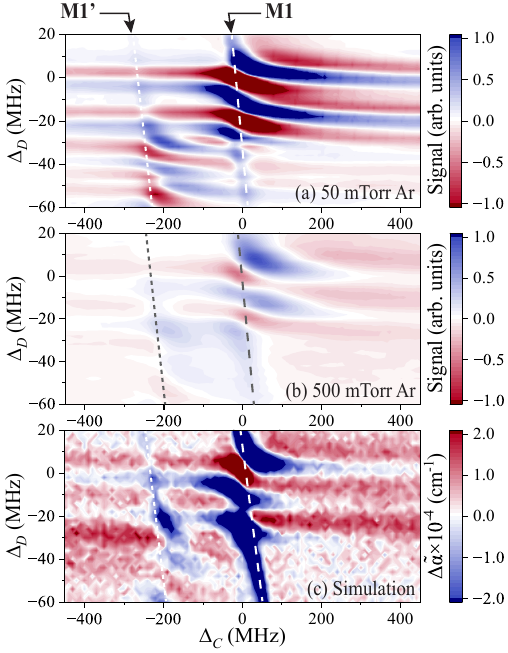}
  \caption{\justifying Absorption maps in the $(+--)$ configuration with Rb cells with nominal Ar pressures of (a) 50~mTorr and (b) 500~mTorr.
  (a) and (b) use the same color scale. (c) QMCWF simulation result including a collisional repumping rate of $3.0 \times 10^7$~s$^{-1}$ for the $5S_{1/2}$ hyperfine states, and hyperfine mixing rates of $3.0 \times 10^7$~s$^{-1}$ and 
  $4.0 \times 10^7$~s$^{-1}$ for the $5P_{1/2}$ and $5D_{3/2}$-levels, respectively.}
  \label{fig:50and500mTorrAr}
\end{figure}

\begin{figure}[htb]
\centering
\includegraphics[width=0.49\textwidth]{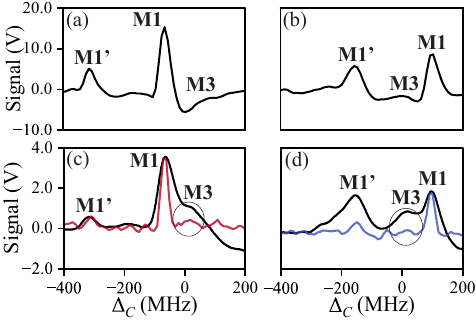}
\caption{\justifying Selected EIT spectra obtained using the vapor cells with added Ar. Laser beams are in the $(+--)$ configuration. Panels (a) and (b) show spectra at 50~mTorr of Ar with $\Delta_D=70$ and $-100~\text{MHz}$, respectively. 
Panels (c) and (d) show data analogous to those in (a) and (b), but for 500~mTorr of Ar. The red and blue curves are simulated plots, scaled to match the experimental data. The M3 signals are indicated by the black ovals. The assumed collisional repumping and %dephasing 
hyperfine mixing rates are the same as in Fig.~\ref{fig:50and500mTorrAr}(c). The number of trajectories per set of parameters $(\Delta_C, v_z)$ is 5000. Note the different $y$-scales in panels (a) and (b) and in panels (c) and (d).}
\label{fig:+--plots}
\end{figure}

Here, we investigate the impact of 50~mTorr and 500~mTorr of Ar on the EIT response in the $(+--)$ configuration, which has a higher EIT signal strength than the $(+-+)$ configuration.
The effects of Ar in  $(+--)$ configuration have not been studied in~\cite{Duspayev2025}.
Figs.~\ref{fig:50and500mTorrAr}(a) and \ref{fig:50and500mTorrAr}(b) show $(+--)$ spectroscopic maps obtained using 4-cm-long vapor cells filled with Rb vapor and 50 and 500~mTorr of Ar, respectively. The parameters of the laser beams are the same as described in Sec.~\ref{sec:excitation scheme}. The Ar gas causes Rydberg line shifts reported at $-350$~MHz/Torr in~\cite{dash2024}. In Figs.~\ref{fig:50and500mTorrAr} and~\ref{fig:+--plots} these shifts are subtracted on the $\Delta_C$-axes, so that the line patterns do not show the net pressure-dependent shift. The overall signal strengths in 50 and 500~mTorr of Ar are reduced by factors of $\sim 2$ and $\sim 10$ compared with those in the 7.5-cm-long Ar-free cell, respectively. The M1 and M1' EIT modes as well as the horizontal striping appear unaltered at 50~mTorr. With 500~mTorr of Ar, they are still observed but weaker and broadened by several tens of MHz in $\Delta_C$. The observed broadening is in general agreement with level-broadening estimates of $\sim 20$~MHz~\cite{dash2024} for Rb Rydberg atoms in 500~mTorr of Ar.
Fig.~\ref{fig:50and500mTorrAr}(c) shows a QMCWF simulation result that includes repumping of the $5S_{1/2}$ hyperfine states by velocity-changing collisions, hyperfine mixing of $5P_{1/2}$, and hyperfine mixing of $5D_{3/2}$ at 
respective rates of $3.0 \times 10^7$~s$^{-1}$,
$3.0 \times 10^7$~s$^{-1}$ and $4.0 \times 10^7$~s$^{-1}$. The Rabi frequencies in the simulation are the same as in Sec.~\ref{subsec:Rb_cell_+--},
and the number of quantum trajectories is 500 
for each set of parameters $(\Delta_C, \Delta_D, v_z)$.  Good qualitative agreement between Figs.~\ref{fig:50and500mTorrAr}(b) and~\ref{fig:50and500mTorrAr}(c) is observed.

Selected EIT spectra are shown in Figs.~\ref{fig:+--plots}(a)-\ref{fig:+--plots}(b) and~\ref{fig:+--plots}(c)-\ref{fig:+--plots}(d) for Ar pressures of 50 and 500~mTorr, respectively. There, the power of the dressing laser was increased to 4~mW for higher SNR, with other beam parameters as described in Sec.~\ref{sec:excitation scheme}. In Figs.~\ref{fig:+--plots}(a) and \ref{fig:+--plots}(c) [\ref{fig:+--plots}(b) and \ref{fig:+--plots}(d)], we set $\Delta_D = $~70~MHz [-100~MHz], both of which are outside the range shown in Fig.~\ref{fig:50and500mTorrAr}.
As seen in Fig.~\ref{fig:+--plots}, the M1 and the M1' peaks become lower and broader with increasing Ar pressure. In Figs.~\ref{fig:+--plots}(c) and \ref{fig:+--plots}(d), a side peak adjacent to M1, denoted as M3, appears. 
The M3 signals occur on the blue (red) side of the M1 signal for positive (negative) $\Delta_D$ and are found at $\Delta_C \approx 0$ (for $\Delta_P=0$). At small $|\Delta_D|$, as in Fig.~\ref{fig:50and500mTorrAr}, the M1 and M3 modes approach each other, eventually coalescing and becoming indiscernible.

The M3 signals are mostly attributed to the hyperfine collisions and level broadening of Rb $5D_{3/2}$.  
To reproduce the M3 signals in the simulations, we apply the same repumping and hyperfine collision rates as those used in the simulation shown in Fig.~\ref{fig:50and500mTorrAr}(c). Moreover, we increase the number of trajectories to 5000 per set of $(\Delta_C, v_z)$-values to improve the SNR of the Monte-Carlo simulation. The simulation data are shown in Figs.~\ref{fig:+--plots}(c) and \ref{fig:+--plots}(d).
To limit computation time, the step size in $\Delta_C$ was set at 10~MHz. The simulated M1 and M1' signals in Fig.~\ref{fig:+--plots} rise well above the statistical noise of the Monte Carlo simulation. The simulated M3 signals, indicated by the black ovals, become borderline visible.

We attribute the M3 signals to a build-up of population in the $5D_{3/2}$-levels, caused by hyperfine-collision-induced level broadening. At $\Delta_C \approx 0$, the incoherently populated $5D_{3/2}$-levels become coupled to relatively long-lived Rydberg levels. The resultant drop in population of $5P_{1/2}$ and $5S_{1/2}$ produces a reduction in absorption on the probe transition, which is experimentally observed as the M3 signal.

\section{Conclusion}
\label{sec:conclusion}
We have investigated EIT modes and collision effects in collinear three-photon Rydberg-EIT in room-temperature Rb vapor cells for two beam-direction configurations. We used a dressed-state analysis as well as numerical studies based on a quantum Monte Carlo trajectory method to explain our experimental observations. 
The quantum trajectory simulations, which utilize Hilbert spaces of up to 108 states, confirm the essential roles of hyperfine structure, optical pumping, radiative decays through the allowed $\vert 5P_J, F', m_{F'} \rangle$ states,
velocity-changing collisions and hyperfine mixing.  
The EIT signal in the $(+--)$ beam-direction configuration 
is narrower and approximately four times stronger than that in the $(+-+)$ configuration.
Therefore, the former will be advantageous in multi-photon Rydberg-EIT applications that require high SNR and narrow linewidths~\cite{Thaicharoen2019,brown2023,bohaichuk2023,allinson2024}.

In the presence of 500~mTorr of Ar gas, velocity- and hyperfine-changing collisions reduce the EIT peak amplitudes and cause line broadening,
and a collision-induced spectral feature is observed. Its emergence is attributed to collisional broadening of the $5D_{3/2}$-levels, which allows incoherent population buildup in these levels.
Our results obtained using Rb cells with 50~mTorr and 500~mTorr of Ar background gas could be useful in future applications of multi-photon Rydberg-EIT in spectroscopy~\cite{allinson2025} and plasma electric-field sensing~\cite{ionsourcepaper, dash2025, vorobiov2025_plasmaeit, andersonAVS2026}.
Our results indicate that Rydberg-EIT in vapor cells may sometimes be affected by velocity-changing collisions caused by spurious gases, especially if EIT beam sizes are large.
Future modeling and simulation work may focus on accurate accounting of Gaussian laser beam profiles, on the roles of collisional fine-structure-, hyperfine-, Zeeman- and Rydberg-state $\ell$-mixing, and on an explicit treatment of velocity-changing collisions.

\section*{Acknowledgments}
\label{sec:acknowledgments}
We would like to thank Dr. Bineet Dash, Xinyan Xiang, Anne Graf, Lefeng Zhou, and Xinyu Feng for technical assistance and helpful discussions. We also thank Dr. Bineet Dash for providing the LabVIEW data-acquisition program. This work was supported by NSF Grant No. PHY-2412535.

\nocite{Dalibard1992,Molmer1993,safronova2004}
\bibliography{bibliography}

@article{Gallagher1968Rubidium,
  title = {Rubidium and Cesium Excitation Transfer in Nearly Adiabatic Collisions with Inert Gases},
  author = {Gallagher, A.},
  journal = {Phys. Rev.},
  volume = {172},
  issue = {1},
  pages = {88--96},
  numpages = {9},
  year = {1968},
  month = {Aug},
  publisher = {American Physical Society},
  doi = {10.1103/PhysRev.172.88},
  url = {https://aps.org}
}

@article{Bulos1971NuclearSpin,
  title = {Nuclear-Spin Inertia and Pressure Broadening of $^{2}{P}_{1/2}$ {H}anle-Effect Signals},
  author = {Bulos, B. R. and Happer, W.},
  journal = {Phys. Rev. A},
  volume = {4},
  issue = {3},
  pages = {849--853},
  numpages = {5},
  year = {1971},
  month = {Sep},
  publisher = {American Physical Society},
  doi = {10.1103/PhysRevA.4.849},
  url = {https://link.aps.org/doi/10.1103/PhysRevA.4.849}
}

@article{Thornton2011,
    title = {Velocity changing collisions in the laser saturation spectra of $^{87}${R}b {$D_2$} {F}"=2},
    journal = {Opt. Comm.},
    volume = {284},
    pages = {2890-2894},
    year = {2011},
    doi = {https://doi.org/10.1016/j.optcom.2011.02.057},
    url = {https://www.sciencedirect.com/science/article/pii/S0030401811002240},
    author = {D. E. Thornton and G. T. Phillips and G. P. Perram}
}

@Article{Dalibard1992,
  author   = {Dalibard, J. and Castin, Y. and M{\"o}lmer, K.},
  journal  = {Phys. Rev. Lett.},
  title    = {Wave-function approach to dissipative processes in quantum optics},
  year     = {1992},
  pages    = {580--583},
  doi      = {10.1103/PhysRevLett.68.580},
  issue    = {5},
  numpages = {0},
  url      = {https://link.aps.org/doi/10.1103/PhysRevLett.68.580},
}

@Article{Molmer1993,
  author    = {K. M{\"o}lmer and Y. Castin and J. Dalibard},
  journal   = {J. Opt. Soc. Am. B},
  title     = {{Monte Carlo} wave-function method in quantum optics},
  year      = {1993},
  pages     = {524--538},
  volume    = {10},
  doi       = {10.1364/JOSAB.10.000524},
  publisher = {Optica Publishing Group},
  url       = {https://opg.optica.org/josab/abstract.cfm?URI=josab-10-3-524},
}

@Article{Xue2019,
  author  = {Xue, Y. and Hao, L. and Jiao, Y. and Han, X. and Bai, S. and Zhao, J. and Raithel, G.},
  journal = {Phys. Rev. A},
  title   = {{{R}ydberg electromagnetically induced transparency in a large Hilbert space}},
  year    = {2019},
  pages   = {053426},
  volume  = {99},
  doi     = {10.1103/PhysRevA.99.053426},
  issue   = {5},
  url     = {https://link.aps.org/doi/10.1103/PhysRevA.99.053426},
}

@article{boller1991,
  title = {Observation of electromagnetically induced transparency},
  author = {Boller, K.-J. and Imamo\ifmmode \breve{g}\else \u{g}\fi{}lu, A. and Harris, S. E.},
  journal = {Phys. Rev. Lett.},
  volume = {66},
  issue = {20},
  pages = {2593--2596},
  numpages = {0},
  year = {1991},
  month = {May},
  publisher = {American Physical Society},
  doi = {10.1103/PhysRevLett.66.2593},
  url = {https://link.aps.org/doi/10.1103/PhysRevLett.66.2593}
}

@article{mohapatra2007,
  title = {Coherent Optical Detection of Highly Excited {R}ydberg States Using Electromagnetically Induced Transparency},
  author = {Mohapatra, A. K. and Jackson, T. R. and Adams, C. S.},
  journal = {Phys. Rev. Lett.},
  volume = {98},
  issue = {11},
  pages = {113003},
  numpages = {4},
  year = {2007},
  month = {Mar},
  publisher = {American Physical Society},
  doi = {10.1103/PhysRevLett.98.113003},
  url = {https://link.aps.org/doi/10.1103/PhysRevLett.98.113003}
}

@ARTICLE{Holloway2014,
  author={C. L. {Holloway} and J. A. {Gordon} and S. {Jefferts} and A. {Schwarzkopf} and D. A. {Anderson} and S. A. {Miller} and N. {Thaicharoen} and G. {Raithel}},
  journal={IEEE Trans. Antennas Propag.}, 
  title={Broadband {R}ydberg Atom-Based Electric-Field Probe for {SI}-Traceable, Self-Calibrated Measurements}, 
  year={2014},
  volume={62},
  number={12},
  pages={6169-6182},
  url={https://ieeexplore.ieee.org/document/6910267}}

@article{Badger_2001,
	doi = {10.1088/0953-4075/34/22/107},
	url = {https://dx.doi.org/10.1088/0953-4075/34/22/107},
	year = {2001},
	month = {nov},
	publisher = {},
	volume = {34},
	number = {22},
	pages = {L749},
	author = {S. D. Badger and I. G. Hughes and C. S. Adams},
	title = {Hyperfine effects in electromagnetically induced transparency},
	journal = {J. Phys. B}
}

@article{Yang2011,
  title = {Electromagnetically induced transparency without a {D}oppler background in a multilevel ladder-type cesium atomic system},
  author = {Yang, B. and Gao, J. and Zhang, T. and Wang, J.},
  journal = {Phys. Rev. A},
  volume = {83},
  pages = {013818},
  numpages = {5},
  year = {2011},
  doi = {10.1103/PhysRevA.83.013818},
  url = {https://link.aps.org/doi/10.1103/PhysRevA.83.013818}
}

@article{he2013,
  title = {Ladder-type electromagnetically induced transparency with optical pumping effect},
  author = {He, Z.-S. and Tsai, J.-H. and Chang, Y.-Y. and Liao, C.-C. and Tsai, C.-C.},
  journal = {Phys. Rev. A},
  volume = {87},
  issue = {3},
  pages = {033402},
  numpages = {7},
  year = {2013},
  month = {Mar},
  publisher = {American Physical Society},
  doi = {10.1103/PhysRevA.87.033402},
  url = {https://link.aps.org/doi/10.1103/PhysRevA.87.033402}
}

@article{Duspayev2025,
  title = {Hyperfine structure and collisions in three-photon {R}ydberg electromagnetically induced transparency},
  author = {Duspayev, A. and Raithel, G.},
  journal = {Phys. Rev. A},
  volume = {111},
  issue = {3},
  pages = {032821},
  numpages = {9},
  year = {2025},
  month = {Mar},
  publisher = {American Physical Society},
  doi = {10.1103/PhysRevA.111.032821},
  url = {https://link.aps.org/doi/10.1103/PhysRevA.111.032821}
}

@article{Kondo:15,
    author = {J. M. Kondo and N. \v{S}ibali\'{c} and A. Guttridge and C. G. Wade and N. R. De Melo and C. S. Adams and K. J. Weatherill},
    journal = {Opt. Lett.},
    number = {23},
    pages = {5570--5573},
    title = {Observation of interference effects via four-photon excitation of highly excited {R}ydberg states in thermal cesium vapor},
    volume = {40},
    month = {Dec},
    year = {2015},
    url = {https://opg.optica.org/ol/abstract.cfm?URI=ol-40-23-5570},
    doi = {10.1364/OL.40.005570}
}

@article{Thaicharoen2024,
  title = {Rydberg electromagnetically induced transparency of ${}^{85}\mathrm{Rb}$ vapor in a cell with {N}e buffer gas},
  author = {Thaicharoen, N. and Cardman, R. and Raithel, G.},
  journal = {Phys. Rev. Appl.},
  volume = {21},
  pages = {064004},
  numpages = {7},
  year = {2024},
  publisher = {American Physical Society},
  doi = {10.1103/PhysRevApplied.21.064004},
  url = {https://link.aps.org/doi/10.1103/PhysRevApplied.21.064004}
}

@article{dash2024,
    author = {Dash, B. and Thaicharoen, N. and Paradis, E. and Duspayev, A. and Raithel, G.},
    title = {Rydberg electromagnetically induced transparency of $^{85}${R}b vapor in {A}r, {N}e, and {N}$_2$ gases},
    journal = {APL Quantum},
    volume = {2},
    number = {1},
    pages = {016132},
    year = {2025},
    month = {03},
    issn = {2835-0103},
    doi = {10.1063/5.0237759},
    url = {https://doi.org/10.1063/5.0237759}
}

@article{sujap2022,
    author = {Su, H.-J. and Liou, J.-Y. and Lin, I-C. and Chen, Y.-H.},
    title = {Optical pumping effects on high-contrast {R}ydberg electromagnetically induced transparency},
    journal = {J. Appl. Phys.},
    volume = {132},
    number = {24},
    pages = {244401},
    year = {2022},
    month = {12},
    issn = {0021-8979},
    doi = {10.1063/5.0127828},
    url = {https://doi.org/10.1063/5.0127828}
}

@article{Thaicharoen2019,
    title = {Electromagnetically induced transparency, absorption, and microwave-field sensing in a {R}b vapor cell with a three-color all-infrared laser system},
    journal = {Phys. Rev. A},
    author = {Thaicharoen, N. and Moore, K.R. and Anderson, D.A. and Powel, R.C. and Peterson, E. and Raithel, G.},
    volume = { 100},
    year = {2019},
    pages = {063427},
    URL = {http://dx.doi.org/10.1103/PhysRevA.100.063427}
}

@article{carr2012,
    author = {Carr, C. and Tanasittikosol, M. and Sargsyan, A. and Sarkisyan, D. and Adams, C. S. and Weatherill, K. J.},
    journal = {Opt. Lett.},
    number = {18},
    pages = {3858--3860},
    publisher = {OSA},
    title = {Three-photon electromagnetically induced transparency using {R}ydberg states},
    volume = {37},
    month = {Sep},
    year = {2012},
    url = {http://opg.optica.org/ol/abstract.cfm?URI=ol-37-18-3858},
    doi = {10.1364/OL.37.003858}
}

@book{gallagher,
  title={Rydberg Atoms},
  author={Gallagher, T.F.},
  year={2005},
  publisher={Cambridge University Press}
}

@article{ma20,
author = {L. Ma and E. Paradis and G. Raithel},
journal = {Opt. Express},
pages = {3676--3685},
title = {{DC} electric fields in electrode-free glass vapor cell by photoillumination},
volume = {28},
year = {2020},
url = {http://www.opticsexpress.org/abstract.cfm?URI=oe-28-3-3676},
doi = {10.1364/OE.380748},
}

@article{brown2023,
  title = {Very-high- and ultrahigh-frequency electric-field detection using high angular momentum {R}ydberg states},
  author = {Brown, R. C. and Kayim, B. and Viray, M. A. and Perry, A. R. and Sawyer, B. C. and Wyllie, R.},
  journal = {Phys. Rev. A},
  volume = {107},
  issue = {5},
  pages = {052605},
  numpages = {8},
  year = {2023},
  month = {May},
  publisher = {American Physical Society},
  doi = {10.1103/PhysRevA.107.052605},
  url = {https://link.aps.org/doi/10.1103/PhysRevA.107.052605}
}

@book{Scullybook,
  address       = {Cambridge},
  title         = {Quantum optics},
  isbn          = {978-0-511-81399-3},
  url           = {http://ebooks.cambridge.org/ref/id/CBO9780511813993},
  publisher     = {Cambridge University Press},
  author        = {Scully, M. O. and Zubairy, M. S.},
  year          = {1997},
  doi           = {10.1017/CBO9780511813993}
}

@article{Lei2025,
  title = {Collisional Broadening of $^{85}${R}b {R}ydberg Levels: Conclusions for Vapor-Cell Manufacture},
  author = {Lei, M. and Eckel, S. P. and Norrgard, E. B. and Prajapati, N. and Artusio-Glimpse, A. B. and Simons, M. T. and Holloway, C. L.},
  journal = {Phys. Rev. Appl.},
  volume = {23},
  number = {3},
  pages = {034028},
  year = {2025},
  publisher = {American Physical Society (APS)},
  doi = {10.1103/PhysRevApplied.23.034028},
  url = {https://doi.org}
}

@article{bohaichuk2023,
  title = {Three-photon {R}ydberg-atom-based radio-frequency sensing scheme with narrow linewidth},
  author = {Bohaichuk, S. M. and Ripka, F. and Venu, V. and Christaller, F. and Liu, C. and Schmidt, M. and K\"ubler, H. and Shaffer, J. P.},
  journal = {Phys. Rev. Appl.},
  volume = {20},
  issue = {6},
  pages = {L061004},
  numpages = {6},
  year = {2023},
  month = {Dec},
  publisher = {American Physical Society},
  doi = {10.1103/PhysRevApplied.20.L061004},
  url = {https://link.aps.org/doi/10.1103/PhysRevApplied.20.L061004}
}

@article{ionsourcepaper,
  title = {Electric Field Analysis in a Cold-Ion Source Using {S}tark Spectroscopy of {R}ydberg Atoms},
  author = {Duspayev, A. and Raithel, G.},
  journal = {Phys. Rev. Appl.},
  volume = {19},
  issue = {4},
  pages = {044051},
  numpages = {11},
  year = {2023},
  month = {Apr},
  publisher = {American Physical Society},
  doi = {10.1103/PhysRevApplied.19.044051},
  url = {https://link.aps.org/doi/10.1103/PhysRevApplied.19.044051}
}

@MISC{SteckRb85,
   author =       "D. A. Steck",
   title =        "Rubidium 85 {D} Line Data",
   year =         "revision 2021",
   url =          "http://steck.us/alkalidata"
 }

@article{jau2020,
  title = {Vapor-Cell-Based Atomic Electrometry for Detection Frequencies below 1 {kHz}},
  author = {Jau, Y.-Y. and Carter, T.},
  journal = {Phys. Rev. Appl.},
  volume = {13},
  issue = {5},
  pages = {054034},
  numpages = {11},
  year = {2020},
  month = {May},
  publisher = {American Physical Society},
  doi = {10.1103/PhysRevApplied.13.054034},
  url = {https://link.aps.org/doi/10.1103/PhysRevApplied.13.054034}
}

@article{safronova2004,
  title = {Relativistic many-body calculations of electric-dipole matrix elements, lifetimes, and polarizabilities in rubidium},
  author = {Safronova, M. S. and Williams, C. J. and Clark, C. W.},
  journal = {Phys. Rev. A},
  volume = {69},
  issue = {2},
  pages = {022509},
  numpages = {8},
  year = {2004},
  month = {Feb},
  publisher = {American Physical Society},
  doi = {10.1103/PhysRevA.69.022509},
  url = {https://link.aps.org/doi/10.1103/PhysRevA.69.022509}
}

@article{prajapati2023highell,
    title={High angular momentum coupling for enhanced {R}ydberg-atom sensing in the very-high frequency band}, 
    author={N. Prajapati and J. W. Kunzler and A. B. Artusio-Glimpse and A. Rotunno and S. Berweger and M. T. Simons and C. L. Holloway and C. M. Gardner and M. S. Mcbeth and R. A. Younts},
    journal = {J. Appl. Phys.},
    volume = {135},
    number = {7},
    pages = {074402},
    year = {2024},
    month = {02},
    issn = {0021-8979},
    doi = {10.1063/5.0179496},
    url = {https://doi.org/10.1063/5.0179496}
}

@article{QMC2018OpticsExpress,
    author = {L. Zhang and S. Bao and H. Zhang and G. Raithel and J. Zhao and L. Xiao and S. Jia},
    journal = {Opt. Express},
    title   = {{Interplay between optical pumping and {R}ydberg {EIT} in magnetic fields}},    
    volume = {26},
    issue = {23},
    pages = {29931-29944},
    numpages = {14},
    year = {2018},
    month = {Nov},
    doi = {10.1364/OE.26.029931},
    url = {https://doi.org/10.1364/OE.26.029931}
}

@article{EITQuantumDefect,
  title = {Measurement of absolute transition frequencies of $^{87}\mathrm{Rb}$ to $\mathit{nS}$ and $\mathit{nD}$ {R}ydberg states by means of electromagnetically induced transparency},
  author = {Mack, M. and Karlewski, F. and Hattermann, H. and H\"ockh, S. and Jessen, F. and Cano, D. and Fort\'agh, J.},
  journal = {Phys. Rev. A},
  volume = {83},
  issue = {5},
  pages = {052515},
  numpages = {8},
  year = {2011},
  month = {May},
  publisher = {American Physical Society},
  doi = {10.1103/PhysRevA.83.052515},
  url = {https://link.aps.org/doi/10.1103/PhysRevA.83.052515}
}

@article{EITStarkEffect,
  title = {Microwave electrometry with {R}ydberg atoms in a vapour cell using bright atomic resonances},
  author = {Sedlacek, J. A. and Schwettmann, A. and Kübler, H. and Löw, R. and Pfau, T. and Shaffer, J. P.},
  journal = {Nat. Phys.},
  volume = {8},
  issue = {11},
  pages = {819-824},
  numpages = {6},
  year = {2012},
  month = {Nov},
  publisher = {Nature},
  doi = {10.1038/nphys2423},
  url = {https://doi.org/10.1038/nphys2423}
}

@article{SWM,
  title = {Coherent Microwave-to-Optical Conversion via Six-Wave Mixing in {R}ydberg Atoms},
  author = {Han, J. and Vogt, T. and Gross, C. and Jaksch, D. and Kiffner, M. and Li, W.},
  journal = {Phys. Rev. Lett.},
  volume = {120},
  issue = {9},
  pages = {093201},
  numpages = {6},
  year = {2018},
  month = {Mar},
  publisher = {American Physical Society},
  doi = {10.1103/PhysRevLett.120.093201},
  url = {https://link.aps.org/doi/10.1103/PhysRevLett.120.093201}
}

@article{EITSWM,
  title = {Slow-Light Six-Wave Mixing at Low Light Intensities},
  author = {Kang, H. and Hernandez, G. and Zhu, Y.},
  journal = {Phys. Rev. Lett.},
  volume = {93},
  issue = {7},
  pages = {073601},
  numpages = {4},
  year = {2004},
  month = {Aug},
  publisher = {American Physical Society},
  doi = {10.1103/PhysRevLett.93.073601},
  url = {https://link.aps.org/doi/10.1103/PhysRevLett.93.073601}
}

@article{EITSinglePhoton2,
author = {Y. O. Dudin  and A. Kuzmich },
title = {Strongly Interacting {R}ydberg Excitations of a Cold Atomic Gas},
journal = {Science},
volume = {336},
number = {6083},
pages = {887-889},
year = {2012},
doi = {10.1126/science.1217901},
URL = {https://www.science.org/doi/abs/10.1126/science.1217901}
}

@article{EITSinglePhoton1,
  title = {Quantum nonlinear optics with single photons enabled by strongly interacting atoms},
  author = {Peyronel, T. and Firstenberg, O. and Liang, Q.-Y. and Hofferberth, S. and Gorshkov, A. V. and Pohl, T. and Lukin, M. D. and Vuletić, V.},
  journal = {Nature},
  volume = {8},
  issue = {7409},
  pages = {57-60},
  numpages = {4},
  year = {2012},
  month = {Aug},
  publisher = {Nature},
  doi = {10.1038/nature11361},
  url = {https://doi.org/10.1038/nature11361}
}

@article{dash2025,
      title={Electric field diagnostics in a continuous rf plasma using {R}ydberg-{EIT}}, 
      author={B. Dash and X. Xiang and D. Feng and E. Paradis and G. Raithel},
      year={2025},
      journal={arXiv:2512.16867},
      url={https://arxiv.org/abs/2512.16867}
}

@article{AngleMatching,
  title = {Doppler- and recoil-free laser excitation of {R}ydberg states via three-photon transitions},
  author = {Ryabtsev, I. I. and Beterov, I. I. and Tretyakov, D. B. and Entin, V. M. and Yakshina, E. A.},
  journal = {Phys. Rev. A},
  volume = {84},
  issue = {5},
  pages = {053409},
  numpages = {6},
  year = {2011},
  month = {Nov},
  publisher = {American Physical Society},
  doi = {10.1103/PhysRevA.84.053409},
  url = {https://link.aps.org/doi/10.1103/PhysRevA.84.053409}
}

@article{Gorshkov2011,
  title = {Photon-Photon Interactions via {R}ydberg Blockade},
  author = {Gorshkov, A. V. and Otterbach, J. and Fleischhauer, M. and Pohl, T. and Lukin, M. D.},
  journal = {Phys. Rev. Lett.},
  volume = {107},
  issue = {13},
  pages = {133602},
  numpages = {5},
  year = {2011},
  month = {Sep},
  publisher = {American Physical Society},
  doi = {10.1103/PhysRevLett.107.133602},
  url = {https://link.aps.org/doi/10.1103/PhysRevLett.107.133602}
}

@article{highellpaper,
  title = {High-angular-momentum {R}ydberg states in a room-temperature vapor cell for dc electric-field sensing},
  author = {Duspayev, A. and Cardman, R. and Anderson, D. A. and Raithel, G.},
  journal = {Phys. Rev. Res.},
  volume = {6},
  issue = {2},
  pages = {023138},
  numpages = {12},
  year = {2024},
  month = {May},
  publisher = {American Physical Society},
  doi = {10.1103/PhysRevResearch.6.023138},
  url = {https://link.aps.org/doi/10.1103/PhysRevResearch.6.023138}
}

@article{allinson2024,
  title = {Simultaneous multiband radio-frequency detection using high-orbital-angular-momentum states in a {R}ydberg-atom receiver},
  author = {Allinson, G. and Jamieson, M. J. and Mackellar, A. R. and Downes, L. and Adams, C. S. and Weatherill, K. J.},
  journal = {Phys. Rev. Res.},
  volume = {6},
  issue = {2},
  pages = {023317},
  numpages = {8},
  year = {2024},
  month = {Jun},
  publisher = {American Physical Society},
  doi = {10.1103/PhysRevResearch.6.023317},
  url = {https://link.aps.org/doi/10.1103/PhysRevResearch.6.023317}
}

@article{patrick2025,
    author = {Patrick, L. and Schlossberger, N. and Hammerland, D. F. and Prajapati, N. and McDonald, T. and Berweger, S. and Talashila, R. and Artusio-Glimpse, A. B. and Holloway, C. L.},
    title = {Imaging of induced surface charge distribution effects in glass vapor cells used for {R}ydberg atom-based sensors},
    journal = {AVS Quantum Sci.},
    volume = {7},
    number = {2},
    pages = {024401},
    year = {2025},
    month = {06},
    issn = {2639-0213},
    doi = {10.1116/5.0264378},
    url = {https://doi.org/10.1116/5.0264378}
}

@article{shieldingpaper,
  title = {Effects of inert background gases and photoillumination on three-color electromagnetically induced transparency of rubidium vapor},
  author = {Duspayev, A. and Dash, B. and Raithel, G.},
  journal = {Phys. Rev. Appl.},
  volume = {24},
  issue = {4},
  pages = {044057},
  numpages = {9},
  year = {2025},
  month = {Oct},
  publisher = {American Physical Society},
  doi = {10.1103/fyf8-4wml},
  url = {https://link.aps.org/doi/10.1103/fyf8-4wml}
}

@article{allegrini2022,
    author = {Allegrini, M. and Arimondo, E. and Orozco, L. A.},
    title = {Survey of Hyperfine Structure Measurements in Alkali Atoms},
    journal = {J. Phys. Chem. Ref. Data},
    volume = {51},
    number = {4},
    pages = {043102},
    year = {2022},
    month = {11},
	issn = {0047-2689},
    doi = {10.1063/5.0098061},
    url = {https://doi.org/10.1063/5.0098061}
}

@article{eitrepumper2021,
    author = {Prajapati, N. and Robinson, A. K. and Berweger, S. and Simons, M. T. and Artusio-Glimpse, A. B. and Holloway, C. L.},
    title = {Enhancement of electromagnetically induced transparency based {R}ydberg-atom electrometry through population repumping},
    journal = {Appl. Phys. Lett.},
    volume = {119},
    number = {21},
    pages = {214001},
    year = {2021},
    month = {11},
	issn = {0003-6951},
    doi = {10.1063/5.0069195},
    url = {https://doi.org/10.1063/5.0069195}
}

@article{OP_Cs,
  title = {Ladder-type electromagnetically induced transparency with optical pumping effect},
  author = {He, Z.-S. and Tsai, J.-H. and Chang, Y.-Y. and Liao, C.-C. and Tsai, C.-C.},
  journal = {Phys. Rev. A},
  volume = {87},
  issue = {3},
  pages = {033402},
  numpages = {7},
  year = {2013},
  month = {Mar},
  publisher = {American Physical Society},
  doi = {10.1103/PhysRevA.87.033402},
  url = {https://link.aps.org/doi/10.1103/PhysRevA.87.033402}
}

@article{DAS2024112340,
    title = {Experimental and theoretical study of magnetic field effect on {EIT} resonances in a multilevel {V}-type atomic system},
    journal = {Chem. Phys.},
    volume = {584},
    pages = {112340},
    year = {2024},
    issn = {0301-0104},
    doi = {https://doi.org/10.1016/j.chemphys.2024.112340},
    url = {https://www.sciencedirect.com/science/article/pii/S0301010424001691},
    author = {C. Das and S. Dey and B. Ray}
}

@article{PhysRevA.73.033803,
  title = {Dark-line atomic resonances in a submicron-thin {R}b vapor layer},
  author = {Sargsyan, A. and Sarkisyan, D. and Papoyan, A.},
  journal = {Phys. Rev. A},
  volume = {73},
  issue = {3},
  pages = {033803},
  numpages = {7},
  year = {2006},
  month = {Mar},
  publisher = {American Physical Society},
  doi = {10.1103/PhysRevA.73.033803},
  url = {https://link.aps.org/doi/10.1103/PhysRevA.73.033803}
}

@article{PhysRevA.65.023814,
  title = {Comparative study of the transient evolution of {H}anle electromagnetically induced transparency and absorption resonances},
  author = {Valente, P. and Failache, H. and Lezama, A.},
  journal = {Phys. Rev. A},
  volume = {65},
  issue = {2},
  pages = {023814},
  numpages = {8},
  year = {2002},
  month = {Jan},
  publisher = {American Physical Society},
  doi = {10.1103/PhysRevA.65.023814},
  url = {https://link.aps.org/doi/10.1103/PhysRevA.65.023814}
}

@article{Das:21,
    author = {B. C. Das and A. Das and D. Bhattacharyya and S. De},
    journal = {J. Opt. Soc. Am. B},
    number = {2},
    pages = {584--594},
    publisher = {Optica Publishing Group},
    title = {Effects of vector magnetic field on electromagnetically induced transparency with lin $\bot$ lin polarization},
    volume = {38},
    month = {Feb},
    year = {2021},
    url = {https://opg.optica.org/josab/abstract.cfm?URI=josab-38-2-584},
    doi = {10.1364/JOSAB.411632}
}

@article{allinson2025,
  title = {Determination of quantum defects and core polarizability of atomic cesium via terahertz and radio-frequency spectroscopy in thermal vapor},
  author = {Allinson, G. and Downes, L. A. and Adams, C. S. and Weatherill, K. J.},
  journal = {Phys. Rev. A},
  volume = {111},
  issue = {6},
  pages = {062802},
  numpages = {12},
  year = {2025},
  month = {Jun},
  publisher = {American Physical Society},
  doi = {10.1103/PhysRevA.111.062802},
  url = {https://link.aps.org/doi/10.1103/PhysRevA.111.062802}
}

@article{Moon:08,
	author = {H. S. Moon and L. Lee and J. B. Kim},
	journal = {Opt. Express},
	number = {16},
	pages = {12163--12170},
	publisher = {Optica Publishing Group},
	title = {Double resonance optical pumping effects in electromagnetically induced transparency},
	volume = {16},
	month = {Aug},
	year = {2008},
	url = {https://opg.optica.org/oe/abstract.cfm?URI=oe-16-16-12163},
	doi = {10.1364/OE.16.012163}
}

@article{Ma_jphyscomm2020,
    doi = {10.1088/2399-6528/abb57c},
    url = {https://doi.org/10.1088/2399-6528/abb57c},
    year = {2020},
    month = {sep},
    publisher = {IOP Publishing},
    volume = {4},
    number = {9},
    pages = {095020},
    author = {Ma, L. and Raithel, G.},
    title = {Doppler narrowing, {Z}eeman and laser beam-shape effects in {$\Lambda$}-type electromagnetically induced transparency on the $^{85}${R}b {D}2 line in a vapor cell},
    journal = {J. Phys. Commun.}
}

@article{tang2025,
  title = {Noise performance of a {R}ydberg atomic receiver},
  author = {Tang, Y. and Zhou, H. and Yang, C. and Ren, S. and Wang, S. and Lu, C.},
  journal = {Phys. Rev. A},
  volume = {112},
  issue = {4},
  pages = {043106},
  numpages = {14},
  year = {2025},
  month = {Oct},
  publisher = {American Physical Society},
  doi = {10.1103/3zlx-g5xl},
  url = {https://link.aps.org/doi/10.1103/3zlx-g5xl}
}

@article{Ma2017,
  title = {Paschen-{B}ack effects and {R}ydberg-state diamagnetism in vapor-cell electromagnetically induced transparency},
  author = {Ma, L. and Anderson, D. A. and Raithel, G.},
  journal = {Phys. Rev. A},
  volume = {95},
  issue = {6},
  pages = {061804(R)},
  numpages = {6},
  year = {2017},
  month = {Jun},
  publisher = {American Physical Society},
  doi = {10.1103/PhysRevA.95.061804},
  url = {https://link.aps.org/doi/10.1103/PhysRevA.95.061804}
}

@article{schlossberger2026,
  title = {Resolving magnetic-sublevel structure in {R}ydberg {A}utler-{T}ownes spectra with arbitrary rf polarization},
  author = {Schlossberger, N. and Talashila, R. and Oliver, S. B. and Prajapati, N. and Watterson, W. J. and Holloway, C. L.},
  journal = {Phys. Rev. A},
  volume = {114},
  issue = {3},
  pages = {033701},
  numpages = {6},
  year = {2026},
  month = {Sep},
  publisher = {American Physical Society},
  doi = {10.1103/w5k9-1mlh},
  url = {https://link.aps.org/doi/10.1103/w5k9-1mlh}
}

@article{vorobiov2025_plasmaeit,
      title={Non-perturbative {2D} spatial measurements of electric fields within a plasma sheath}, 
      author={M. Vorobiov and R. Behary and W. Torg and N. DeStefano and S. Mordijck and E. Thomas Jr. and S. C. Thakur and C. T. Fancher and N. Malvania and S. Aubin and E. E. Mikhailov and I. Novikova},
      year={2025},
      journal={arXiv:2511.12322},
      url={https://arxiv.org/abs/2511.12322}      
}

@article{andersonAVS2026,
    author = {Anderson, D. A. and Duspayev, A. and Raithel, G. and Simons, M. and Holloway, C. L.},
    title = {Non-invasive measurements of plasma electric fields using {R}ydberg electromagnetically induced transparency},
    journal = {AVS Quantum Sci.},
    volume = {8},
    number = {3},
    pages = {034404},
    year = {2026},
    month = {09},
	issn = {2639-0213},
    doi = {10.1116/5.0340257},
    url = {https://doi.org/10.1116/5.0340257}
}

@article{nez1993,
    title = {Optical frequency determination of the hyperfine components of the $5{S}_{1/2}$-$5{D}_{3/2}$ two-photon transitions in rubidium},
    journal = {Opt. Comm.},
    volume = {102},
    pages = {432-438},
    year = {1993},
    doi = {https://doi.org/10.1016/0030-4018(93)90417-4},
    url = {https://www.sciencedirect.com/science/article/pii/0030401893904174},
    author = {F. Nez and F. Biraben and R. Felder and Y. Millerioux},
}

@misc{SM,
  note = {See Supplemental Material at
  [URL will be inserted by publisher] for details of the quantum Monte Carlo wave-function model, including the Hilbert-space construction, optical couplings, decay channels, and numerical implementation, which includes
  Refs.~\cite{Dalibard1992, Molmer1993, safronova2004}}
}

\end{document}

\title{Supplemental Material} 
% move into main text later
\maketitle

\section{QMCWF Model}

The EIT spectral maps shown in the main text are reproduced in their essential parts by a simulation of the time-dependent Lindblad equation. The three EIT laser fields in the simulation are linearly polarized, as in the experiment. The relevant $^{85}$Rb Hilbert state space consists of all hyperfine levels $\{ \vert 5S_{1/2}, F, m_F \rangle \}$ $\oplus$ $\{ \vert 5P_{1/2}, F', m_{F'} \rangle \}$ $\oplus$ $\{ \vert 5D_{3/2}, F'', m_{F''} \rangle \}$ $\oplus$ $\{ \vert 25F_{5/2}, F''', m_{F'''} \rangle \}$ $\oplus$ $\{ \vert 5P_{3/2}, \tilde{F}, m_{\tilde{F}} \rangle \}$. The dimension of this Hilbert space is 108. The states are coupled via the coherent optical drives and spontaneous decays. The states among the first four groups are optically coupled by the EIT probe, dressing and coupler lasers as well as through spontaneous decay. The last group (the $5P_{3/2}$ states) is included because they are coupled via spontaneous decay from the $5D_{3/2}$ levels. The spontaneous and thermal decay of the $25F_{5/2}$ Rydberg states is slow in comparison with typical atom-field interaction times. The latter range around $1~\mu$s due to the EIT beam diameters and the thermal velocities of the atoms. In comparison, for the Rydberg state $25F_{5/2}$ used in the experiment we calculate a radiative lifetime of 9.0~$\mu$s at room temperature, with the majority of the decays occurring into the $4D_{3/2}$ and $5D_{3/2}$ levels. Since the radiative decay rates of $25F_{5/2}$ are small, they are not important; nevertheless, radiative decays of $25F_{5/2}$ into $5D_{3/2}$ are included in our model.

Due to the large number of 108 states, we solve the time-dependent Lindblad equation using the quantum trajectory method, which yields an approximation to the density matrix using stochastic wave-functions, or quantum trajectories~\cite{Dalibard1992, Molmer1993, QMC2018OpticsExpress, Xue2019}. For given sets of laser-drive parameters, the EIT-coupler detuning, $\Delta_C$, is typically stepped in steps of 10~MHz, and the dressing detuning, $\Delta_D$, in steps of 2~MHz, over respective ranges from $-450$~MHz to $450$~MHz and from $-60$MHz to $20$~MHz. The coupler detuning is measured relative to the unperturbed transition from $\vert 5D_{3/2}, F''=4 \rangle$ to $\vert25F_{5/2}, F'''=5\rangle$. The dressing detuning is measured relative to the $\vert 5P_{1/2}, F'=3\rangle$ to $\vert5D_{3/2}, F''=4\rangle$. The EIT-probe detuning relative to $\vert5S_{1/2}, F=3\rangle$ to $\vert5P_{1/2}, F'=3\rangle$ transition is held fixed at $\Delta_P = 0$ (as is done in the experiment). While the Rydberg hyperfine structure is not spectroscopically resolved, it is included in the simulation because it affects the overall atom 
distribution over the multi-level system. 

The atoms have three statistically independent velocity components that each follow a one-dimensional Maxwell velocity distribution. 
The longitudinal velocity is sampled in steps of $\Delta v_z=$1~m/s over a range from $-600$~m/s to $600$~m/s,  which approximately equals $3.5 \sigma_v$ with  $\sigma_v = \sqrt{kT/m} \approx 171$~m/s. Since $v_z$ determines the optical Doppler shifts of the coherent optical drives, each value of $v_z$ requires a separate set of quantum trajectory simulations for each of the sampled detunings $(\Delta_C, \Delta_D)$. 
In the compilation of results, the $v_z$-samples are assigned their proper statistical weights, $ [\Delta {v_z} /(\sigma_v \sqrt{2 \pi})] \, \, \exp[  - (v_z/\sigma_v)^2/2]$.

For each set $(\Delta_C, \Delta_D, v_z)$, an ensemble of quantum trajectories, $\{ \vert \psi_i(t=0) \rangle$, $i=1,2 ...i_{\rm{max}}\}$, is initialized in random superpositions of the magnetic sublevels of the upper $^{85}$Rb hyperfine ground state, $|5S_{1/2}, F=3 \rangle$. We use between $i_{\rm{max}} =500$ and 5000 quantum trajectories per set. The transverse velocity, $v_\perp$, is statistically independent of the longitudinal velocity and determines the atom-field interaction time, $t_{\rm{int}}$. The transverse velocity distribution, $P_{v_\perp} (v_\perp)$, corresponds with two spatial degrees of freedom and therefore follows 
$P_{v_\perp} (v_\perp) = \frac{1}{\sigma_v^2} v_{\perp} \exp[  - (v_\perp/\sigma_v)^2/2]$. The resultant probability distribution for $t_{\rm{int}}$, $P_t(t_{\rm{int}})$, is calculated once for the given beam radius, $R$, and $P_{v_\perp} (v_\perp)$. The atom-field interaction times, $t_{{\rm{int}},i}$, for the members of the quantum-trajectory ensemble are then randomly drawn according to the distribution $P_t(t_{\rm{int}})$.

The coherent optical drives have Rabi frequencies, in units Hz, of
\[ \Omega_{n,L, J, F, m_{F}}^{n',L', J', F', m_{F'}} = \frac{-|e| a_0 E_0}{h} \, R 
_{n,L,J}^{n',L', J'} w_{L, J, F, m_{F}}^{L', J', F', m_{F'}}
\quad, \]
with radial matrix elements $R_{n,L, J}^{n',L', J'}$ and angular 
matrix elements $w_{L, J, F, m_{F}}^{L', J', F', m_{F'}}$. There, unprimed and primed variables belong to the lower and upper levels for the three types of coherent EIT drives involved (the probe, dressing and coupling laser beams). 
For the radial matrix elements, defined as $R_{n,L, J}^{n',L', J'} = \int R_{n,L,J} (r) R_{n',L',J'} r^3 d^3r$ for valence-electron wavefunctions $R_{n,L,J}(r)$ and $R_{n',L',J'}(r)$, we use $R_{5S_{1/2}}^{5P_{1/2}} = 5.1842$,  $R_{5S_{1/2}}^{5P_{3/2}} = 5.1776$~\cite{SteckRb85}, $R_{5P_{1/2}}^{5D_{3/2}} = 1.399$, $R_{5P_{3/2}}^{5D_{3/2}}=1.524$~\cite{safronova2004}, and $R_{5D_{3/2}}^{25F_{5/2}}=0.2181$~[in-house calculation], in atomic units and after conversions that account for different formulations of the Wigner-Eckart theorem. The $E_0$ denote the electric-field amplitudes of the respective laser beams. Here, the angular matrix elements are for $\pi$-polarized drive fields, with the quantization axis pointing along $x$ in the main text. The EIT laser drives only couple levels of equal $m_F$.
The optical Rabi frequencies are assumed to be fixed and equal to their values at the beam centers.

The partial spontaneous decay rates from upper state $\vert n',L', J', F', m_{F'} \rangle$ into lower state $\vert n,L, J, F, m_{F} \rangle$ are given by 
\[ \gamma_{n',L', J', F', m_{F'}}^{n,L, J, F, m_{F}} = \frac{4 \alpha \omega^3}{3 c^2} \, (a_0 R 
_{n',L', J'}^{n,L, J} w_{L', J', F', m_{F'}}^{L, J, F, m_{F}})^2
\quad, \]
with the fine structure constant $\alpha$ and transition angular frequencies $\omega$. Primed and unprimed variables belong to the upper and lower levels of the decays. Summing over the lower-level $m_F$-values results in the same decay rate from $(n', L', J', F')$ into $(n, L, J, F)$ for any $m_{F'}$,
\[ \tilde{\gamma}_{n',L', J', F'}^{n,L, J, F} = 
\sum_{m_F} \gamma_{n',L', J', F', m_{F'}}^{n,L, J, F, m_{F}}
\quad.\]
Summing next over the lower-state hyperfine quantum numbers $F$, one obtains the decay rates from the upper fine structure levels, $n'L'_{J'}$, into the allowed lower fine structure levels, $nL_J$,
\[ \Gamma_{n',L', J'}^{n,L, J} = 
\sum_{F}
\tilde{\gamma}_{n',L', J', F'}^{n,L, J, F}
\quad. \]
The $\Gamma_{n',L', J'}^{n,L, J}$ do not depend on hyperfine quantum numbers. Finally, the total decay rates of the fine-structure levels $n'L'_{J'}$ are
\[ \tilde{\Gamma}_{n',L', J'} = \sum_{n,L,J}  \Gamma_{n',L', J'}^{n,L, J}
\quad. \]
The above summations are subject to electric-dipole selection rules.

The effective Hamiltonian used to propagate the quantum trajectories $\vert \psi_i (t) \rangle$ then is
\[ \hat{H} = \frac{h}{2} \sum_{\lambda',\lambda} \vert \lambda' \rangle  \Omega_\lambda^{\lambda'} \langle \lambda \vert + h \sum_{\lambda'} \vert \lambda' \rangle \Big[ \Delta_{\lambda'}-\frac{{\rm{i}} \tilde{\Gamma}_{\lambda'}}{4 \pi} \Big] \langle \lambda' \vert \quad,\] 
where we use the shorthand $\vert \lambda \rangle  = \vert  n,L, J, F, m_{F} \rangle $. Note Rabi frequencies are in Hz and decay rates in s$^{-1}$.  While there are 108 levels, the first sum is only sparsely occupied due to optical selection rules. The second, diagonal term includes dressed-state energies $h \Delta_{\lambda'}$, which depend on laser detunings and hyperfine splittings in Hz, and level decay terms $- {\rm{i}} \hbar \tilde{\Gamma}_{\lambda'}/2$ with the above-defined $\tilde{\Gamma}_{n', L', J'}$ that result in a monotonic decay of the wave-function norm.    

For each parameter set $(\Delta_C, \Delta_D, v_z)$, each of the quantum trajectories in $\{ \vert \psi_i (t) \rangle \}$ is propagated over its randomly drawn interaction time, $t_{{\rm{int}},i}$. The dynamics of the $\vert \psi_i(t) \rangle$ are governed by alternating segments of Hamiltonian evolution and discrete quantum jumps, which model decay processes (radiative decay, collision-induced repumping and hyperfine mixing). A quantum jump occurs when the norm of $\vert \psi_i (t) \rangle$ drops below a random number, which is refreshed after each jump. The quantum jump of the trajectory $\vert \psi_i (t) \rangle$ is  instantaneous and results in a re-set of the $\vert \psi_i (t) \rangle$ into a normalized state. The latter is used as initial state for next segment of Hamiltonian evolution of the trajectory $\vert \psi_i (t) \rangle$. Therefore, the trajectory follows an alternating sequence of Hamiltonian evolutions and discrete quantum jumps. The trajectory is propagated until its interaction time, $t_{{\rm{int}},i}$, is reached. The procedure is repeated for all $\vert \psi_i (t) \rangle$.

Each quantum jump involves several random numbers to select decay modes and to simulate a decay-specific outcome for $\vert \psi_i (t) \rangle$ after the jump. First, the $\vert \psi_i (t) \rangle$ immediately before the jump, the above-defined
decay rates 
$\tilde{\gamma}_{n',L', J', F'}^{n,L, J, F}$, and a new random number are used to determine out of which upper manifold of states, $\vert n', L', J', F'\rangle$, the decay occurs, and into which lower manifold of states, $\vert n, L, J, F\rangle$, the decay leads. Then, $\vert \psi_i (t) \rangle$  is projected into a coherent superposition of upper-level states $\vert n', L', J', F', m_{F'} \rangle$ within the drawn upper-level manifold $\vert n', L', J', F'\rangle$ out of which the decay occurs. This projection, which maintains Zeeman coherences within the drawn upper-level manifold, yields an upper-level-only $\vert \psi_{1,i} (t) \rangle$. Application of spherical electric-dipole jump operators $\hat{{\bf{r}}}^{(1)}_q$ on
$\vert \psi_{1,i} (t) \rangle$ for the emission of $\pi$ ($q=0$), $\sigma^+$ ($q=-1$) and  $\sigma^-$ ($q=1$) photons then yields three
possible states $\vert \psi_{2,q,i} (t) \rangle$ after decay into the drawn lower manifold of states, $\vert n,L,J,F\rangle$. The norms of these three states and another new random number are used to select the polarization of the emitted photon. The thusly selected state $\vert \psi_{2,q,i} (t) \rangle$ is normalized and used as initial state for the next segment of Hamiltonian evolution. 

It is noted that the quantum jumps simulate quantum measurements of photons of well-defined color and polarization. The upper- and lower-state quantum numbers, $(n',L',J',F')$ and $(n,L,J,F)$, are associated with the color of the measured photons. The polarizations of the measured photons are associated with the selected $q$-values. The quantum-jump rules for hyperfine mixing and ground-state repumping due to velocity-changing collisions are fairly straightforward relative to those for the radiative decays.  

The present three-photon EIT scheme involves a combination of single-, double- and triple-resonant excitation channels, which all contribute to the observed absorption signal for the EIT probe laser. There are two methods to determine the Beer's absorption coefficient of the atomic vapor for the EIT probe light. 

In the first method, we count the number of quantum jumps that signify the emission of photons on the $5P_J$ to $5S_{1/2}$ channels, as any photon emitted on any of these channels is preceded by the absorption of an EIT-probe photon from $\vert 5S_{1/2}, F=3\rangle$ into 
$\vert 5P_{1/2}, F=2\rangle$ or into $\vert 5P_{1/2}, F=3\rangle$. Also, atoms exiting in any state other than $5S_{1/2}$ are counted into the net absorption, as atoms exiting in any such state require an EIT probe photon to become excited in the first place. Averaging over the Maxwell distribution and the ensembles of quantum trajectories, we obtain the net single-atom absorption rate within the atom-field interaction region, $\gamma_{1}$. The Beer's absorption coefficient of the vapor for the EIT probe then is
\[\bar{\alpha} (\Delta_C, \Delta_D)= \gamma_1 (\Delta_C, \Delta_D) \frac{n_{\rm{V}} \hbar \omega_P }{I_P}\quad,\]
with atom volume density, $n_{\rm{V}}$, and probe laser intensity, $I_P = c \epsilon_0 E_P^2/2$. The probe laser amplitude, $E_P = h \Omega_{5S_{1/2}}^{5P_{1/2}} / (-|e| a_0 R_{5S_{1/2}}^{5P_{1/2}})$, follows from the radial probe Rabi  frequency in Hz, $\Omega_{5S_{1/2}}^{5P_{1/2}}$, and the probe radial dipole moment, $R_{5S_{1/2}}^{5P_{1/2}} = 5.184$ (in atomic units) entered into the simulation. It is $\Omega_{5S_{1/2}}^{5P_{1/2}} = -|e| a_0 E_P R_{5S_{1/2}}^{5P_{1/2}} /h$.

The absorption 
can also be obtained via a second method as follows.   
For each set of parameters ($\Delta_D$, $\Delta_C$, $v_z$), the density operator is approximated by 
\[ \hat{\rho} (\Delta_C, \Delta_D, v_z) = \Big \langle \sum_{i=1}^{i_{\rm{max}}} \frac{\vert \psi_i(t) \rangle \langle \psi_i(t) \vert}  {\langle \psi_i(t)  \vert \psi_i(t) \rangle } \Big \rangle_{{\rm{time}}}\quad.\]
For our $\pi$-polarized EIT probe, the partial atomic absorption coefficient over the velocity step $\Delta {v_z}$ is
\begin{widetext}
\begin{equation*}
\Delta {\alpha} (\Delta_C, \Delta_D, v_z)  = 
\frac {4 \pi \, n_{\rm{V}}} {h \epsilon_0 \lambda_P}
\sum_{F', m_{F}} 
\frac {
 P _{v_z}(v_z)\,(d_{F=3,m_F}^{F', m_{F}})^2 \,{\rm{Im}}[\rho_{F', m_F}^{  F=3,m_F}]} {\Omega_{F=3,m_F}^{F',m_{F}}} \Delta v_{\rm{z}} \quad.
\label{eq:alpha}
\end{equation*}
\end{widetext}
There, $P_{v_z}$ is the one-dimensional Maxwell velocity distribution, $\lambda_P = 795$~nm is the probe-laser wavelength, 
$\rho_{F', m_F}^{  F=3,m_F}$ is the density matrix element $\langle 5S_{1/2}, F=3, m_F \vert \hat{\rho}(\Delta_C, \Delta_D, v_z) \vert 5P_{1/2}, F',m_{F}  \rangle$, 
$\Omega_{F=3,m_F}^{F',m_{F}}$ is the Rabi frequency (in $\mathrm{Hz}$) of the probe transition between those states for our $\pi$-polarized probe, and $d_{F=3,m_F}^{F',m_{F}}$ is the corresponding transition dipole moment in $\mathrm{C} \mathrm{m}$. To arrive at the net Beer's absorption for the EIT probe, we integrate $ \Delta {\alpha} (\Delta_C, \Delta_D, v_z)$ over the Maxwell velocity distribution.
This yields the EIT spectrum ${\alpha} (\Delta_C, \Delta_D)$. We find $\alpha (\Delta_C, \Delta_D) \approx \bar{\alpha} (\Delta_C, \Delta_D)$ throughout, as expected. Minor deviations between the two methods are on the percent-level and are caused by the statistical nature of the QMCWF method and by finite time-stepping effects. In the latter regard, we note that for the time-step size of the simulation we have chosen 10~ns, which is about one-third of the fastest decay time in the system (which is the $5P_{3/2}$ decay time). This choice represents a compromise between computation speed and accuracy. The photon scattering rate $\gamma_1$ is systematically underestimated by up to a few percent. 

The main parameters of the QMCWF simulation are the radial Rabi frequencies, $\Omega_{5S_{1/2}}^{5P_{1/2}}$, 
$\Omega_{5P_{1/2}}^{5D_{3/2}}$, 
and $\Omega_{5D_{3/2}}^{25F_{5/2}}$, the radius of the EIT probe beam, $R$, and the atom density, $n_{\rm{V}}$, which follows from the cell temperature $T=300$~K~\cite{SteckRb85}. In our simulations we have used
$\Omega_{5S_{1/2}}^{5P_{1/2}} = 25$~MHz, 
$\Omega_{5P_{1/2}}^{5D_{3/2}} = 80$~MHz, $\Omega_{5D_{3/2}}^{25F_{5/2}} = 20$~MHz, and $R
=40~\mu$m. These parameters are qualitatively analogous to those used in the experiment. The radii of the EIT dressing and coupler beams are larger than that of the probe beam and are therefore unimportant. It is noted that due to the statistical nature of the QMCWF method the simulation results carry statistical noise, which is equivalent to photon counting noise in an experiment performed on a relatively small number of atoms. The statistical noise of the QMCWF simulations is evident in the simulation plots shown in the main text. 

\bibliography{bibliography}